\documentclass[trackchanges,twocolumn]{aastex7}

\usepackage{amsmath}
\usepackage{xcolor}

\begin{document}

\title{Jet Power, Bulk Lorentz Factor, Black Hole Spin, and Magnetic Field of Accretion Disk in Jetted Active Galactic Nuclei: A Large Gamma-Ray Emission Sample}

\author[0000-0002-6809-9575]{Dingrong Xiong}
\altaffiliation{Corresponding authors}
\affiliation{Yunnan Observatories, Chinese Academy of Sciences, 396 Yangfangwang, Guandu District, Kunming, 650216, People's Republic of China; \textcolor{blue}{xiongdingrong@ynao.ac.cn}; \textcolor{blue}{baijinming@ynao.ac.cn}}
\affiliation{Key Laboratory for the Structure and Evolution of Celestial Objects, Chinese Academy of Sciences, 396 Yangfangwang, Guandu District, Kunming, 650216, People's Republic of China}
\email{xiongdingrong@ynao.ac.cn}  

\author[0000-0002-5929-0968]{Junhui Fan}
\affiliation{Center for Astrophysics, Guangzhou University, Guangzhou 510006, People's Republic of China}
\affiliation{Astronomy Science and Technology Research Laboratory of Department of Education of Guangdong Province, Guangzhou 510006, People's Republic of China}
\email{fjh@gzhu.edu.cn} 

\author[0000-0003-3564-6437]{Feng Yuan}
\affiliation{Center for Astronomy and Astrophysics and Department of Physics, Fudan University,
Shanghai 200438, People's Republic of China}
\email{fyuan@fudan.edu.cn} 

\author[0000-0002-4419-6434]{Jun-Xian Wang}
\affiliation{Department of Astronomy, University of Science and Technology of China, Hefei 230026, People's Republic of China}
\affiliation{School of Astronomy and Space Science, University of Science and Technology of China, Hefei 230026, People's Republic of China}
\email{jxw@ustc.edu.cn} 

\author[0000-0002-4455-6946]{Minfeng Gu}
\affiliation{Shanghai Astronomical Observatory, Chinese Academy of Sciences, 80 Nandan Road, Shanghai 200030, People's Republic of China}
\email{gumf@shao.ac.cn} 

\author[0000-0002-1935-8104]{Yongquan Xue}
\affiliation{Department of Astronomy, University of Science and Technology of China, Hefei 230026, People's Republic of China}
\affiliation{School of Astronomy and Space Science, University of Science and Technology of China, Hefei 230026, People's Republic of China}
\email{xuey@ustc.edu.cn}

\author[0000-0002-7077-7195]{Jirong Mao}
\affiliation{Yunnan Observatories, Chinese Academy of Sciences, 396 Yangfangwang, Guandu District, Kunming, 650216, People's Republic of China; \textcolor{blue}{xiongdingrong@ynao.ac.cn}; \textcolor{blue}{baijinming@ynao.ac.cn}}
\affiliation{Key Laboratory for the Structure and Evolution of Celestial Objects, Chinese Academy of Sciences, 396 Yangfangwang, Guandu District, Kunming, 650216, People's Republic of China}
\email{jirongmao@mail.ynao.ac.cn} 

\author[0000-0002-1908-0536]{Liang Chen}
\affiliation{Shanghai Astronomical Observatory, Chinese Academy of Sciences, 80 Nandan Road, Shanghai 200030, People's Republic of China}
\email{chenliang@shao.ac.cn} 

\author[0000-0003-1721-151X]{Rui Xue}
\affiliation{Department of Physics, Zhejiang Normal University, Jinhua 321004, People's Republic of China}
\email{ruixue@zjnu.edu.cn} 

\author[0000-0002-1908-0536]{Xu-Liang Fan}
\affiliation{School of Mathematics, Physics and Statistics, Shanghai University of Engineering Science, Shanghai 201620, People's Republic of China}
\email{fanxl@sues.edu.cn} 

\author[0000-0001-5895-0189]{Yongyun Chen}
\affiliation{College of Physics and Electronic Engineering, Qujing Normal University, Qujing 655011, People's Republic of China}
\email{ynkmcyy@yeah.net} 

\author[0000-0003-1028-8733]{Nan Ding}
\affiliation{School of Physical Science and Technology, Kunming University 650214, People's Republic of China}
\email{orient.dn@foxmail.com} 

\author[0000-0002-7072-2522]{Fei Guo}
\affiliation{Yunnan Normal University, Kunming 650500, People's Republic of China}
\email{guofei@ynnu.edu.cn} 

\author[0000-0001-5136-8110]{Jia-Wen Li}
\affiliation{Department of Astronomy, School of Physics and Astronomy, Key Laboratory of Astroparticle Physics of Yunnan Province, Yunnan University, Kunming 650091}
\email{jwliynu@ynu.edu.cn} 

\author[0000-0003-4895-1406]{Dahai Yan}
\affiliation{Department of Astronomy, School of Physics and Astronomy, Key Laboratory of Astroparticle Physics of Yunnan Province, Yunnan University, Kunming 650091}
\email{yandahai@ynu.edu.cn} 

\author[0000-0003-0170-9065]{Y. G. Zheng}
\altaffiliation{Corresponding authors}
\affiliation{Department of Physics, Yunnan Normal University, Kunming, Yunnan, 650092, People's Republic of China; \textcolor{blue}{ynzyg@ynu.edu.cn}}
\affiliation{Key Laboratory of Colleges and Universities in Yunnan Province for High-energy Astrophysics, Kunming, Yunnan, 650500, People's Republic of China}
\email{ynzyg@ynu.edu.cn}

\author{Jinming Bai}
\altaffiliation{Corresponding authors}
\affiliation{Yunnan Observatories, Chinese Academy of Sciences, 396 Yangfangwang, Guandu District, Kunming, 650216, People's Republic of China; \textcolor{blue}{xiongdingrong@ynao.ac.cn}; \textcolor{blue}{baijinming@ynao.ac.cn}}
\affiliation{Key Laboratory for the Structure and Evolution of Celestial Objects, Chinese Academy of Sciences, 396 Yangfangwang, Guandu District, Kunming, 650216, People's Republic of China}
\email{baijinming@ynao.ac.cn}


\begin{abstract}
We present a catalog of physical parameters for powerful jet–accretion disk–black hole systems in one of the largest samples of gamma-ray emitting jetted active galactic nuclei (AGNs), including jet kinetic and radiative powers, jet radiative efficiencies, bulk Lorentz factors, black hole spins, accretion-disk magnetic fields and Compton dominance. Comparing jet kinetic power estimators for blazars, values derived from spectral energy distribution (SED) fitting tend to exceed those estimated via cavity power and other scaling relations. For radiatively efficient AGNs, most sources are inferred to possess high spins; for radiatively inefficient AGNs, many potentially have high spins, though some may differ. This indicates that black hole spin does not effectively distinguish radiatively efficient from inefficient jetted AGNs. Our results suggest accretion-disk magnetic field strength as a key discriminator, proposing a tentative dividing value of $\approx 10^{3.9}$ Gauss between radiatively efficient and inefficient populations. Jet power and bulk Lorentz factor exhibit significant correlations with black hole mass in radiatively efficient AGNs, while weak‑to‑moderate correlations are observed in radiatively inefficient AGNs within narrow accretion‑rate bins. Our analysis reveals that jet power correlates with both disk luminosity and magnetic field strength. Furthermore, correlations linking Eddington ratio and Compton dominance with jet properties are consistent with the jet–accretion connection. Finally, jet radiative power and bulk Lorentz factor show a potential dependence on black hole spin. These results are consistent with the scenario in which jets are powered and accelerated by energy extraction from rapidly spinning black holes via accretion‑disk magnetic fields.
\end{abstract}

\keywords{\uat{Active galactic nuclei}{16} --- \uat{Blazars}{164} --- \uat{Galaxy jets}{601} --- \uat{Galaxy accretion disks}{562} --- \uat{Gamma-ray sources}{633} --- \uat{Black hole physics}{159}}

\section{Introduction}

Jetted active galactic nuclei (AGNs), traditionally known as radio-loud AGNs, represent a unique subset of AGNs, whose black hole-accretion disk systems launch powerful, collimated jets (or outflows of plasma) that extend from sub-parsec to megaparsec scales and transport energy, momentum, and magnetic flux into the host galaxies and intergalactic medium \citep[e.g.,][]{Begelman1984,Padovani2017}. These jets are prominent across the electromagnetic spectrum, generating nonthermal emission from radio wavelengths to TeV gamma-rays through mechanisms such as synchrotron radiation, inverse Compton scattering, and hadronic processes \citep[e.g.,][]{Ghisellini2010,Dermer2012,Yan2014,IceCube2018,Xue2019,Zhang2012,Zheng2017,Abdo2009,2022PhRvD.106j3021X,2024ApJS..271...10W,2025ApJ...990..170X}. Jetted AGNs encompass a variety of subclasses, including blazars (flat-spectrum radio quasars and BL Lacertae objects, i.e., FSRQ and BL Lac), as well as radio galaxies (RDG; encompassing Fanaroff–Riley types I and II, i.e., FRI and FRII), steep-spectrum radio quasars (SSRQ), jetted narrow-line Seyfert 1 galaxies (NLSy1), compact steep-spectrum sources (CSS), and other AGNs exhibiting jets \citep[e.g.,][]{Urry1995,Falomo2014,Padovani2017,Zhang2020}.

Jet power and bulk Lorentz factor are among the most fundamental physical parameters that characterize jets. Measurement of the kinetic power of extragalactic jets is crucial to determine the AGN kinetic luminosity function and its evolution, which informs studies of jet feedback and galaxy evolution, and to assess the contribution of black hole spin to jet production \citep[e.g.,][]{Croton2006,Godfrey2013,Daly2009}. However, accurate measurement of jet kinetic power remains a challenge. Various methods have been employed to estimate it, such as X-ray cavity analysis, hotspots and strong shock approaches, and spectral energy distribution (SED) fitting. Scaling relationships between jet kinetic power and radio or gamma ray luminosity have been established, allowing estimates of jet power from radio or gamma-ray luminosities \citep[e.g.,][]{Willott1999,Cavagnolo2010,Godfrey2013,Daly2019,Dea2009,Ghisellini2014,Nemmen2012,Meyer2011}. \citet{Godfrey2013} applied the hotspot method to estimate jet power for FR II sources and compared the resulting jet power versus radio luminosity relation with that for FR I derived from X-ray cavity measurements, finding approximate agreement between the relations determined separately for FR I and FR II. For beamed AGNs, extended radio luminosity is often used to estimate jet power to minimize the influence of Doppler beaming \citep[e.g.,][]{Nemmen2012,Meyer2011}. Alternatively, radio or high energy gamma-ray luminosities, corrected for beaming effects, can also provide an estimate of jet power in beamed AGNs \citep{Fan2018,Nemmen2012,Xiong2014}.

The bulk Lorentz factor $\Gamma$ characterizes the jet velocity, defined as $1/\sqrt{1-\beta^2}$ in which $\beta$ is the velocity of the jet in units of the speed of light. $\Gamma$ can be derived from several methods: the homogeneous synchrotron self-Compton (SSC) model \citep[e.g.,][]{Marscher1987,Ghisellini1993}, the inhomogeneous relativistic jet model \citep[e.g.,][]{Blandford1979,Chai2012}, and estimates based on variability Doppler factors combined with apparent jet speeds \citep[e.g,][]{Pushkarev2009}. To obtain the beaming factor $f_ {\rm b}$, \citet{Nemmen2012} derived the relation between the observed gamma ray luminosity and $\Gamma$; using this relation, $\Gamma$ can then be estimated. \cite{Liodakis2018} evaluated a number of these methods and found that the variability Doppler factor method is the most accurate. \citet{Chai2012} found a significant correlation between the black hole mass and $\Gamma$, but no connection with the Eddington ratio. This suggests that the Blandford–Znajek mechanism may be the primary process driving jet acceleration. Similarly, \citet{Xiong2014b} reported comparable findings for the Fermi FSRQ sample. However, the direct correlation between black hole spin and $\Gamma$ was not examined.

Black holes are nature's simplest objects, characterized only by their mass, spin (angular momentum), and electric charge - the latter effectively neutralized in realistic astrophysical environments. The spin of a black hole plays a crucial role in many astrophysical processes \citep{Reynolds2019}. The formation of supermassive black holes (SMBHs) in the early Universe remains poorly understood, and whether these SMBHs are rapidly spinning can discriminate between formation scenarios driven by coherent disk accretion and those dominated by chaotic accretion or the merger of smaller black holes \citep{Reynolds2019,Volonteri2005}. 
The technique most widely used to measure black hole spins in AGNs is known as ``X-ray reflection", while a second major method is ``thermal continuum fitting" \citep[e.g.,][]{Reynolds2021,Reynolds2019,Reynolds2014}. In AGNs, more than two dozen SMBHs have relatively robust spin measurements using the above methods \citep{Reynolds2019}. The ``outflow method'' has been proposed and applied to large AGN samples \citep{Daly2016,Daly2019,Daly2011,Daly2024,Daly2014}. One way to assess the outflow method’s reliability is to test consistency with independent spin estimates. For 21 SMBHs with comparisons, spins from the outflow method agree very well with those from continuum fitting and X-ray reflection; for the stellar-mass black hole GX 339-4, the outflow method also shows excellent agreement with the X-ray reflection method \citep{Daly2019,Daly2024,Azadi2023}.

Magnetic fields of accretion disks facilitate angular momentum transport, drive jet launching and collimation, and influence disk stability and heating processes \citep[e.g.,][]{Balbus1991,Balbus1998,Blandford1982,Blandford1977}. The polarized horizon-scale Event Horizon Telescope (EHT) images of M87* and Sgr A* suggest that strong magnetic fields may be a common feature of all black holes \citep{Event2021,Event2024}. In addition to the outflow-based spin estimate, \citet{Daly2019} proposed a method to infer the total accretion‑disk magnetic field strength in radio‑loud AGNs from the accretion rate and black hole mass.

Gamma-ray-emitting (or Fermi-detected) jetted AGNs typically exhibit more pronounced beaming effects and possess more powerful jets compared to non-gamma-ray-emitting AGNs \citep[e.g.,][]{Pushkarev2009, Linford2011, Liodakis2018,Pushkarev2012, Xiong2015}. Therefore, gamma-ray-emitting jetted AGNs serve as an optimal sample for investigating physical properties of powerful jet–accretion disk–black hole systems. 
In Paper I \citep{Xiong2014}, we analyzed the intrinsic gamma-ray luminosity, black hole mass, jet and accretion in Fermi blazars based on data from the Fermi Second LAT AGN Catalog (2LAC). Following the release of the Fermi fourth LAT AGN catalog \citep[4LAC-DR3;][]{Ajello2022}, the number of blazars in the high-latitude 4LAC-DR3 sample increases by more than a factor of three compared with 2LAC, and the catalog also contains a substantial increase in non-blazar AGNs. Estimates of jet power, bulk Lorentz factor, black hole spin, and accretion-disk magnetic fields in the 4LAC-DR3 sample remain poorly constrained and underexplored due to beaming effect, model dependence, methodological limitations, etc. Our new catalog is not free from model dependence or methodological limitations.

In this study (Paper II), we analyze the high‑latitude 4LAC‑DR3 sample, deriving jet power, bulk Lorentz factor, black hole spin, accretion disk magnetic field strength, and jet radiative efficiency. Compared with Paper I, this work: (1) uses a much larger sample including non‑blazars; (2) separates sources into radiatively efficient and inefficient classes; and (3) adds parameters not included in Paper I—black hole spin, accretion‑disk magnetic field strength, Compton dominance, and jet radiative efficiency. A catalog of key parameters for powerful jet–accretion disk–black hole systems is presented, constituting one of the largest jetted AGNs samples with such measurements. We then compare the jet kinetic powers derived from different methods, examine the distributions of these parameters across various jetted AGN types, and analyze inter‑parameter correlations. Based on these results, the general physical properties of powerful jet–accretion disk–black hole systems are characterized for the large sample of gamma-ray emitting jetted AGNs.

This paper is organized as follows: Section 2 describes the sample and methods; Section 3 presents the main results; Section 4 provides discussions; Section 5 summarizes the conclusions. Throughout this work, a concordance $\Lambda$CDM cosmology with $H_0=71~\mathrm{km~s^{-1}~Mpc^{-1}}$, $\Omega_{\mathrm{m}}=0.27$, and $\Omega_{\Lambda}=0.73$ \citep{Spergel2007} is adopted. Energy spectral indices are defined by $f_{\nu}\propto\nu^{-\alpha}$.

\section{Sample and methods} \label{sec:sample}

The high-latitude 4LAC-DR3 AGNs were adopted as the parent sample, which is made up of 3407 gamma-ray emission AGNs, including 755 FSRQs, 1379 BL Lacs, 1208 blazars of uncertain type (BCUs), and 65 non-blazar AGNs \citep{Ajello2022}. To estimate jet power from radio luminosity, counterpart positions of the gamma-ray AGNs were cross-matched with positional coordinates from radio bands at 151 MHz \citep[TGSS;][]{Intema2017}, 1.4 GHz \citep[FIRST;][]{Helfand2015ApJ}, and 15 GHz \citep[OVRO;][]{Richards2011}. \cite{Fan2019} examined the impact of different matching radii on the number of matched sources and identified 5 arcseconds as an optimal radius to associate 3LAC sources with the 151 MHz radio sample. Consequently, the same matching radius was applied for cross-matching the 4LAC sources with the radio catalogs. For sources lacking redshift measurements, the mean redshift of the corresponding subclass was adopted as a substitute. After cross-matching with the 151 MHz radio sample, our dataset comprises 2018 blazars, 15 RDG, 6 NLSY1, 5 CSS, 2 SSRQ, and 4 unclassified jetted AGNs (Table 1). Among these, 729 AGNs (see Tables 2–3) have both black hole mass and accretion-disk luminosity, which are essential for estimating black hole spin and the accretion-disk magnetic field.

The proportions of sources with measured redshift among blazar subclasses are $14\%$ (89/658) for BCU, $63\%$ (447/715) for BL Lac, and $100\%$ (645/645) for FSRQ. All other non-blazar jetted AGNs have measured redshift values, except for radio galaxies and unclassified jetted AGNs, each of which lacks 1 redshift measurement. In the subsequent correlation analysis, 7 radiatively efficient AGNs and 22 radiatively inefficient AGNs lack redshift measurements, most of which are BCU.

\paragraph{Caveats}
Using average redshift values for BCU and BL Lac lacking redshift measurements may introduce additional uncertainty. Spectral features from the host galaxy are easier to detect in nearby objects, so sources without redshifts are likely at above‑average redshift; consequently, adopting an average redshift probably underestimates their true jet power, bulk Lorentz factor, black hole spin, and accretion‑disk magnetic field. When computing mean values of these parameters, the true averages for BCUs and BL Lacs are likely higher than those reported in Section 3.2. For comparisons of jet power across methods, adopting an average redshift does not affect relative results because each method uses the same redshift for a given source, and the comparison relies on relative rather than absolute values (see Fig. 3). In the subsequent correlation analysis, parameters for the few sources missing redshifts are computed using the mean redshift; because such sources are scarce in the correlation sample ($4\%$; 29/729), this does not significantly affect the results.

We repeated all analyses (Sections 2.1-3.3) using only sources with measured redshifts (1184 AGNs); the results remained consistent with those obtained assuming average redshifts.

\subsection{Jet kinetic power} \label{subsec:jet_k_power}

\subsubsection{Jet cavity power} \label{subsec:jet_c_power}

\cite{Cavagnolo2010} determined jet cavity powers across a range of systems from giant elliptical galaxies to brightest cluster galaxies and derived the following relation between jet cavity power and low-frequency radio luminosity:
\begin{equation}
\begin{split} 
\mathrm{log}\,P_{\mathrm{cav}} = 0.64(\pm0.09)\left(\mathrm{log}\,L_{\rm 200-400MHz}-40\right)\\ + 43.54(\pm0.12)~(\mathrm{erg~s^{-1}}).
\end{split} 
\end{equation}
The uncertainty in $P_\mathrm{cav}$ is approximately 0.7 dex.

\citet{Meyer2011} estimated the jet kinetic power of blazars and radio galaxies using the low-frequency extended luminosity at 300 MHz, assuming that $P_\mathrm{kin} = P_\mathrm{cav}$. Following \citet{Meyer2011}, \citet{Nemmen2012} and \citet{Xiong2014} utilized the 300 MHz extended luminosity to estimate the jet kinetic power of gamma-ray emitting blazars. The relation of \citet{Cavagnolo2010} spans nearly six orders of magnitude in jet cavity power and thus encompasses the jet powers of gamma-ray emitting AGNs; accordingly, we adopted the extended 300 MHz luminosity to estimate jet kinetic power for these sources.
\begin{figure}
    \centering
    \includegraphics[width=0.52\textwidth]{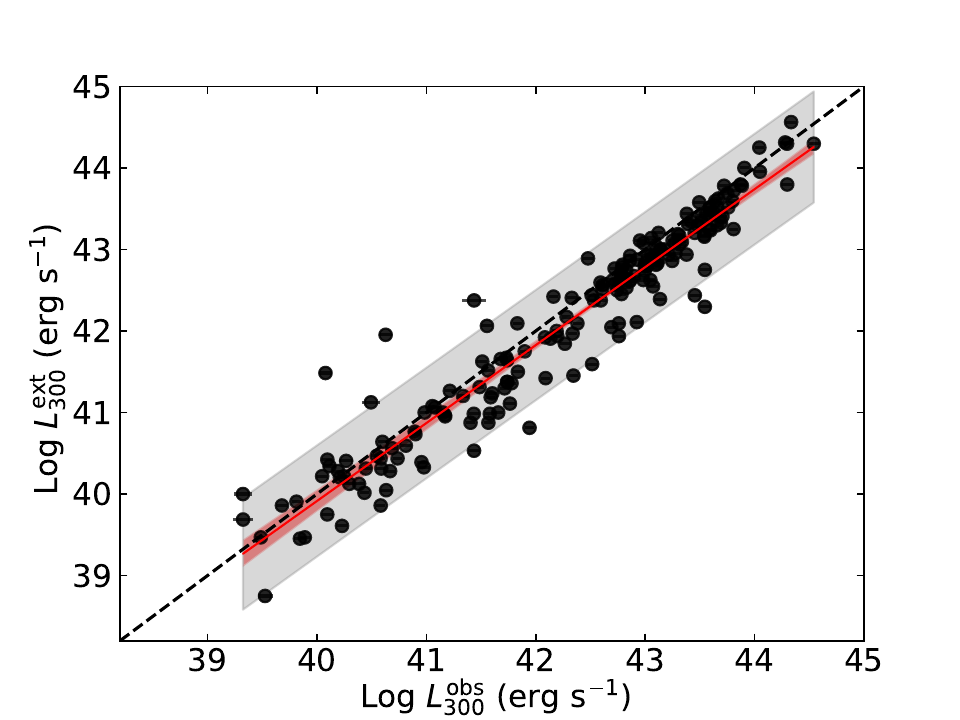}
    \caption{Extended 300 MHz radio luminosity versus observed 300 MHz radio luminosity for beamed blazars. The y‑axis data are from \citet{Nemmen2012}, while the x‑axis values are the observed 151 MHz fluxes extrapolated to 300 MHz. Solid circles denote data points; the dashed line marks equality of extended and observed luminosities. The red solid line represents the best-fit linear model obtained through ordinary least squares fitting; the red shaded area is the 95\% confidence band and the gray area the 95\% prediction band. The correlation coefficient for this fit is $r = 0.96$, which corresponds to a confidence level of 22.1$\sigma$. For several sources the extended 300 MHz radio luminosity appears to exceed the observed luminosity, likely due to extrapolating from 151 MHz with a fixed spectral index and the non‑simultaneity of the extended and observed measurements.}
\end{figure}

Low-frequency total radio emission in beamed blazars comprises both core and extended components \citep{Fan2018,Massaro2013}. Extended radio emission is required to mitigate beaming effects. The extended 300 MHz radio luminosities for the sample were collected directly from Table S1 of \citet{Nemmen2012}. They estimated jet kinetic power for blazars with extended radio emission observed by the VLA by measuring the extended (core-subtracted) flux from VLA images (typically at 1.4 GHz) and extrapolating this flux to obtain the extended 300 MHz radio luminosity.

Most blazars in the sample lacked directly measured extended radio luminosities; therefore, we estimated the extended luminosity from the observed radio luminosity as follows. The observed 151 MHz flux was converted to 300 MHz assuming a spectral index $\alpha=0.57$ \citep{Giroletti2016}. The energy flux (or $\nu F_{\rm \nu}$) was K-corrected according to 
\begin{equation}
S_{\rm \nu} = S^{\rm obs}_{\rm \nu}(1 + z)^{\alpha - 1}.
\end{equation}
The luminosity was then calculated using the relation 
\begin{equation}
L = 4\pi d_L^2 S_{\rm \nu},
\end{equation}
where $d_L$ is the luminosity distance. 

Fig. 1 illustrates a strong correlation between the extended and observed radio luminosities for objects with extended 300 MHz radio luminosities, showing that the core component constitutes a larger proportion at higher luminosities; however, at lower luminosities (less than $10^{41}~\mathrm{erg\,s^{-1}})$, the extended luminosity is nearly equal to the observed luminosity. The extended 300 MHz radio data are from \citet{Nemmen2012}, while the observed values are the measured 151 MHz fluxes extrapolated to 300 MHz. The fitted relationship can be expressed as
\begin{equation}
\mathrm{log}\,L_{\rm 300}^{\rm ext}=0.956(\pm0.02){\rm log}\,L_{\rm 300}^{\rm obs}+1.663(\pm0.833).
\end{equation}
The standard deviation (1$\sigma$) of residuals between the best-fit linear model and the data (presented in Fig. 1) is approximately 0.3 dex. Combining the standard deviation of residuals with the uncertainty in $P_\mathrm{cav}$, the uncertainty in $P_\mathrm{kin}$ for blazars without directly measured extended radio luminosities was estimated to be $\approx 0.76$ dex. 9.37\% (189/2018) of the blazar sample had extended fluxes taken from \citet{Nemmen2012}, corresponding to an uncertainty in $P_{\rm kin}$ of $\approx 0.7$ dex. Equation (4) was used to derive extended luminosities for the remaining blazars. A strong correlation exists between extended and observed 300 MHz radio luminosities (22.1$\sigma$ significance); the uncertainty from this scaling relation is included in the jet power estimates, so using it to infer extended luminosities for the remaining sources is justified.

The spectral index for NLSy1 was also assumed to be $\alpha = 0.57$, since their physical properties resemble those of blazars \citep[e.g.][]{Zhou2003,Foschini2011,Foschini2015}, and the jet kinetic power was similarly calculated using equations (1)-(4). For radio galaxies and other jetted AGNs, the spectral index was set to $\alpha = 0.8$ \citep{Cavagnolo2010}, while for CSS and SSRQ, an extended-lobe spectral index of $\alpha = 1.0$ was adopted \citep{Urry1995,Xiong2015}. For non‑beamed jetted AGNs, the extended radio component was expected to dominate the low‑frequency radio emission. Therefore, after converting the 151 MHz flux to 300 MHz, the jet kinetic power was calculated directly using equation (1).

\subsubsection{Jet kinetic powers estimated by other methods} \label{subsec:jet_other_power}

In addition to the jet cavity power, this subsection presented estimates of jet kinetic power obtained through other methods.

For blazars, jet kinetic powers were estimated using equations (5) and (6), with $L_{\rm 151}/2.1$ employed to account for the combined core and extended radio emission \citep{Fan2018,Fan2019}. In equations (7) and (9), intrinsic gamma‑ray or radio luminosities serve as inputs, where intrinsic luminosity $L_{\rm in} = f_{\rm b} L_{\rm obs}$, with the beaming factor $f_{\rm b} = 1 - \cos(1/\Gamma)$, and $\Gamma$ representing the bulk Lorentz factor. The estimation of $\Gamma$ can be found in subsection 2.3.

\paragraph{Model-dependent estimation of jet kinetic power}
Due to the absence of reliable empirical approaches for directly measuring the kinetic energy of AGN jets, researchers have extensively relied on a model-based proxy for jet kinetic power established by \citet{Willott1999}, which integrates synchrotron minimum energy estimates with the self-similar evolution model of radio galaxies. \citet{Willott1999} presented a formulation that links the jet kinetic power to the radio luminosity at 151 MHz, expressed as follows:
\begin{equation}
    P_{\rm W} \approx f^{3/2} \times 1.7 \times 10^{45} \left( \frac{L_{151}}{10^{44} \, \text{erg s}^{-1} } \right)^{6/7} \, \text{erg s}^{-1},
    \label{eq:sample_label}
\end{equation}
where $f$ is a parameter representing systematic uncertainties in the model assumptions, with $1 \leq f \leq 20$ \citep{Willott1999,Godfrey2013}.

\paragraph{Method of hotspots}
\cite{Godfrey2013} proposed a new method to estimate the jet kinetic power based on measurements of the observed size and luminosity of the jet terminal hotspots, deriving the relation: 
\begin{equation}
\begin{aligned}
    P_{\rm Hot} = g \times (1.5\pm0.5) \times 10^{44}\\ \times \left( \frac{L_{151}}{1.9 \times 10^{41} \, \text{erg s}^{-1} } \right)^{0.67\pm0.05} \, \text{erg s}^{-1},
\end{aligned}
\end{equation}
where the factor $g$ is set to 2.

\paragraph{Estimating jet kinetic power via gamma-ray luminosity}
\citet{Nemmen2012} found that jets from AGNs and GRBs follow the same relation between jet cavity power and intrinsic 100 MeV to 100 GeV gamma-ray luminosity, expressed as:
\begin{equation}
\begin{aligned}
\mathrm{log}\,P_{\mathrm{Gamma-ray}} = 0.98(\pm0.02)\mathrm{log}\,L_{\mathrm{\gamma}}+1.6(\pm0.9).
\end{aligned}
\end{equation}

\paragraph{Jet kinetic power estimation via the Blandford and K\"onigl model}

\citet{Foschini2024} applied the simplified Blandford and K\"onigl model to estimate the jet kinetic power; adopting the model's typical parameter values and treating $k2$ as a constant yields the following simplified relation:
\begin{equation}
   P_{\mathrm{BK}}= \left( 3.9 \times 10^{44} \right) \left( \frac{S_{\rm 15GHz} d_{\rm L,9}^2}{1+z} \right)^{\frac{12}{17}}\, \text{erg s}^{-1},
\end{equation}
where the radio flux density $S_{\rm 15GHz}$ is measured in Jy and the luminosity distance $d_{\rm L,9}$ is in Gpc.

When deriving equation (8), \cite{Foschini2024} did not account for beaming effect; therefore, no beaming correction was applied when using equation (8) to calculate jet kinetic power. The radio flux data at 1.4 GHz and 15 GHz were incomplete, with 1866 radio flux measurements obtained at 1.4 GHz and 434 at 15 GHz. Additionally, 146 SED-fitted jet kinetic powers were obtained directly from \citet{Ghisellini2014}.

\paragraph{Method of the fundamental line/plane of black hole}
The``strong‑shock" method was applied to calculate the jet kinetic (or beam) power of FR II radio sources \citep{Daly2016,Daly2019}. 
To compute jet power for a large sample in a model-independent way, combining the black‑hole fundamental line/plane yields an empirical expression for the jet kinetic power as follows:
\begin{equation}
\mathrm{log}\,P_{\mathrm{FL}} = 0.75(\pm0.07)\mathrm{log}\,L_{\rm 1.4GHz}\\ + 14.93(\pm2.61)~(\mathrm{erg~s^{-1}}).
\end{equation}
When computing the 1.4 GHz radio luminosity, a spectral index of $\alpha = 0.0$ was assumed.

In addition to the methods above, a common approach to estimate jet kinetic power is SED fitting, from which the jet kinetic power is computed as $P_{\rm SED}=P_{\rm p}+P_{\rm e}+P_{\rm B}$, where $P_{\rm p}$ is the power in cold protons, $P_{\rm e}$ in radiating electrons, and $P_{\rm B}$ the Poynting flux \citep[e.g.,][]{Ghisellini2010}. SED-fitting jet kinetic powers were taken directly from \citet{Ghisellini2014} for the subsequent comparative analysis.

We compared the jet kinetic powers derived with these different methods and present the results in Section 3.1.

\subsection{Jet radiative power} \label{subsec:jet_r_power}

Jet radiative power can be estimated from the bolometric jet luminosity \citep{Ghisellini2012}, typically via
\begin{equation}
    P_{\rm r} = \pi R^2 \Gamma^2 \beta c \frac{L'_{\text{bol}}}{4 \pi R^2 c} = \frac{\Gamma^2 L_{\text{bol}}^{\text{obs}}}{4 \delta^4},
    \label{eq:placeholder}
\end{equation}
where $L'_{\rm bol}$ and $L_{\rm bol}^{\rm obs}$ are the intrinsic and observed bolometric jet luminosities, respectively, and $\Gamma$ and $\delta$ denote the bulk Lorentz factor and Doppler factor, respectively. For blazars, with $\sin \theta_v \sim \frac{1}{\Gamma}$ and $\Gamma \approx\delta$, the radiative jet power is approximated as 
\begin{equation}
P_{\rm r}\sim \frac{L_{\text{bol}}^{\text{obs}}}{4 \Gamma^2}.
\end{equation}
The sum of the two peak luminosities can be estimated as \citep{Nemmen2012}
\begin{equation}
    L^{\text{LL}}_{\text{bol}} = L^{\text{obs}}_{\text{IC}} + L^{\text{obs}}_{\text{syn}},
\end{equation}
where $L^{\text{obs}}_{\text{IC}}$ and $L^{\text{obs}}_{\text{syn}}$ represent the luminosities of the inverse Compton and synchronous peaks in the SED, respectively. Due to the broad-band coverage of the two peaks in the SED, the integrated luminosity of each peak exceeds the corresponding peak luminosity. The $L^{\text{LL}}_{\text{bol}}$ from equation (12) should be defined as a lower limit of bolometric jet luminosity. Based on quasi‑simultaneous SEDs, the bolometric jet luminosity (integrated over all bands) is on average 8.3 times the sum of the two peak luminosities \citep{Xue2016}:
\begin{equation}
    L^{\text{obs}}_{\text{bol}} \sim 8.3\times L^{\text{LL}}_{\text{bol}}.
\end{equation}

We extracted $\nu F_{\rm \nu}$ at the high‑ and low‑energy peaks from 4LAC-DR3 to derive $L^{\text{obs}}_{\text{IC}}$ and $L^{\text{obs}}_{\text{syn}}$. The $\Gamma$ and $\delta$ are described in subsection 2.3. For non‑blazar jetted AGNs, when calculating $P_r$ using equation (10), the average $\Gamma$ and $\delta$ values of the corresponding subclass (5 RDG and 3 NLSy1) were adopted in cases where direct measurements were not available. A significant caveat: the RDG and NLSy1 samples are small. After excluding RDG and NLSy1 for which averages were used, the trends or correlations related to radiative power remained consistent. The uncertainty in $P_{\rm r}$ was calculated by propagating the errors associated with $L^{\text{obs}}_{\text{bol}}$, $\Gamma$ and $\delta$.

$63.2\%$ (1276/2018) of blazars had $P_{\rm r}$ computed directly via equation (11), while the remaining $36.8\%$ had $P_{\rm r}$ estimated using the scaling relation of equation (14). All 6 NLSy1 in the sample were calculated using equation (10); however, due to missing $L^{\text{obs}}_{\text{IC}}$ and/or $L^{\text{obs}}_{\text{syn}}$ measurements, $P_{\rm r}$ could not be obtained for many other types of jetted AGNs (7 RDG, 5 CSS, 2 SSRQ, 4 unclassified jetted AGNs). The strong correlation between $L_{\rm \gamma}^{\rm int}$ and $P_{\rm r}$ is illustrated in Fig 2.
\begin{figure}
    \centering
    \includegraphics[width=0.52\textwidth]{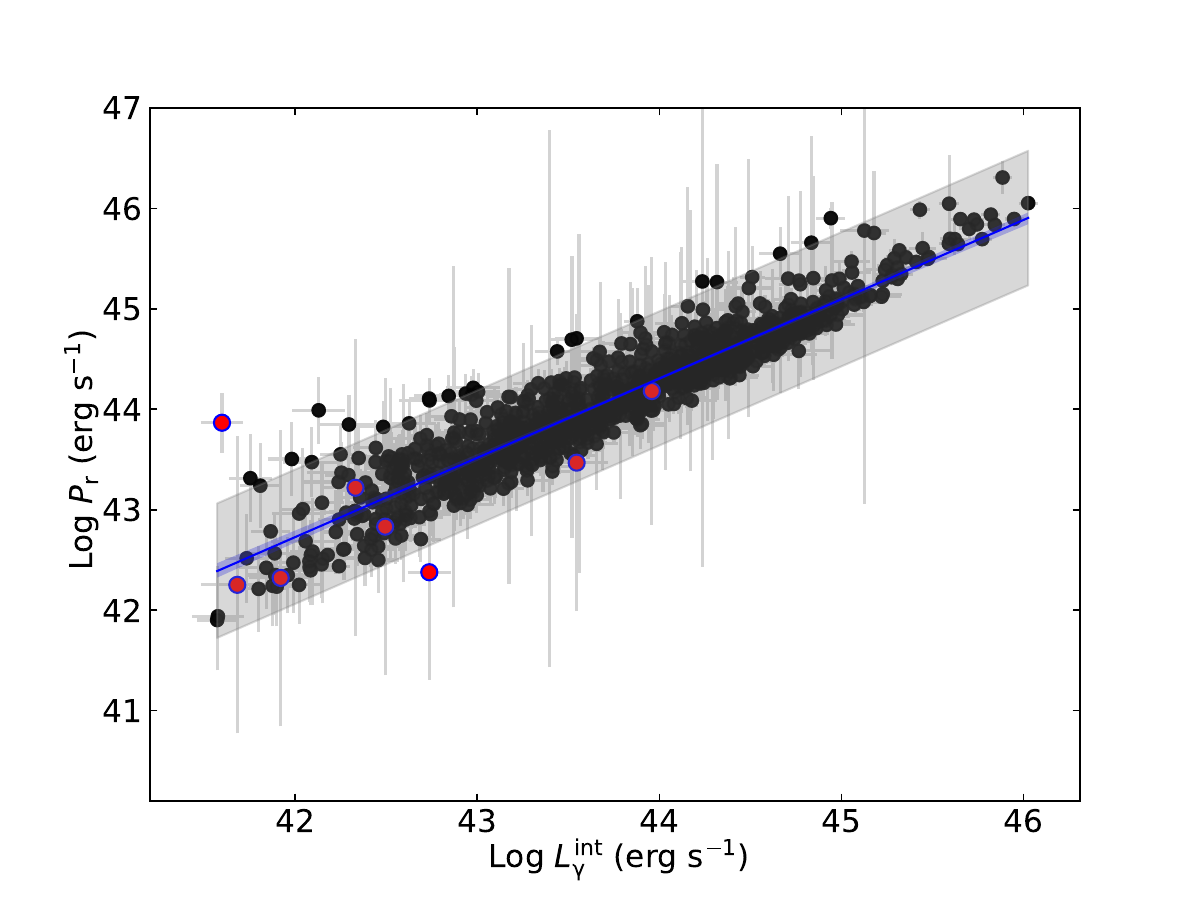}
    \caption{The relationship between jet radiative power and intrinsic gamma-ray luminosity, with black circles representing blazars and red circles indicating radio galaxies. The blue solid line shows the best-fit linear model obtained via ordinary least squares fitting; the blue shaded area corresponds to the 2$\sigma$ confidence band, while the gray region depicts the 2$\sigma$ prediction band. The correlation coefficient for this fit is $r = 0.94$ (20$\sigma$ significance), with a scatter of 0.05 dex.}
\end{figure}
This relation is given by:
\begin{equation}
\mathrm{log}\,P_{\rm r}=0.788(\pm0.008){\rm log}\,L_{\rm \gamma}^{\rm int}+9.63(\pm0.34).
\end{equation}
For jetted AGNs lacking direct $L^{\text{obs}}_{\text{IC}}$ and/or $L^{\text{obs}}_{\text{syn}}$ measurements, the jet radiative power was estimated using equation (14); the corresponding uncertainty combined the standard deviation of the fit residuals with the mean error of directly measured $P_{r}$, i.e., $\sqrt{\sigma^2+<{\rm err}>^2}$. For blazars, the uncertainty in jet power estimated via the scaling relation (equation 14) was 0.359 dex (0.307 dex from the fit residual standard deviation and 0.19 dex from the mean error of directly measured $P_{r}$), whereas for other AGN types it was 1.3 dex (0.307 dex from the fit residual standard deviation and 1.26 dex from the mean error of directly measured $P_{r}$). A strong, highly significant correlation exists between jet radiative power and intrinsic gamma‑ray luminosity; accounting for the fit uncertainty when estimating the radiative power error. Therefore using this scaling relation to estimate jet radiative power is justified. However, the uncertainty for other AGN types appears to be greater than the differences observed between the classes, preventing any conclusive comparisons among the different AGN classes.

The jet radiative power provides a lower limit on the total jet power, since if all kinetic and Poynting power were converted into radiation, the jet would cease \citep{Ghisellini2012}. 

The total jet power is the sum of its kinetic and radiative powers with $P_\mathrm{kin} = P_\mathrm{cav}$, expressed as $P_{\rm total}=P_{\rm kin}+P_{\rm r}$, while the jet radiative efficiency is defined as $\eta_{\rm rad}=P_{\rm r}/P_{\rm total}$.

\subsection{Bulk Lorentz factor} \label{subsec:bulk}

\cite{Liodakis2018} estimated the variability $\delta_{\rm var}$ and $\Gamma_{\rm var}$ for a large AGN sample. The $\delta_{\rm var}$ was derived from the ratio of the variability brightness temperature to the equipartition brightness temperature, and $\Gamma_{\rm var}$ was calculated from the apparent speed of the jet together with $\delta_{\rm var}$.

The $\Gamma$ and $\delta$ in our sample (181 AGNs) were adopted from \citet{Liodakis2018}. For blazars not listed in \citet{Liodakis2018}, the empirical beaming factor in the radio band -- observed gamma ray luminosity relation ($f_{\rm b}-L_{\rm \gamma}^{\rm obs}$) of \citet{Xiong2025} was used to estimate $\Gamma$. To derive the empirical relation, beaming factors were obtained from radio data, and then used gamma ray luminosity to estimate the radio beaming factor for sources lacking values from \citet{Liodakis2018}. \citet{Nemmen2012} derived a relation between observed gamma ray luminosity and the beaming factor for the 2LAC blazar sample; for sources lacking measured beaming factors, $f_{\rm b}$ was estimated from the observed gamma ray luminosity using this scaling relation, and \citet{Xiong2025} found a consistent scaling relation based on 4LAC data. The average ratio between the $\Gamma$ values from \cite{Liodakis2018} and those derived from the scaling relation for the same sources is $1.35 \pm 0.1$. A clear caveat: the scaling gives systematically lower $\Gamma$ values.

The uncertainty in $\Gamma$ is assumed to be comparable to that in $\delta$ for blazars. For blazars with direct measurements, the uncertainty in $\Gamma$ was a factor of 0.59, while for those without direct measurements the mean uncertainty was approximately a factor of 1.2 \citep{Xiong2025}. For non-blazar jetted AGNs, when parameters were calculated using the mean values of $\Gamma$ and $\delta$, both the mean uncertainty of direct measurements and the standard error of the mean were taken into account.

\begin{deluxetable*}{ccccccccl}
\tabletypesize{\scriptsize}
\tablenum{1}
\tablewidth{0pt}
\tablecaption{Jet power, bulk lorentz factor and radiative efficiency in jetted active galactic nuclei. \label{tab:1}} 
\tablehead{
\colhead{4FGL name} & \colhead{Class} &\colhead{$z$} &\colhead{${\rm Log}~P_{\rm kin}$}  &  \colhead{$\Gamma$} &\colhead{${\rm Log}~P_{\rm r}$} & \colhead{${\rm Log}~P_{\rm total}$}  &  \colhead{$\eta_{\rm rad}$} &  \colhead{Name of Counterpart}\\
\colhead{} & \colhead{} &  \colhead{} &\colhead{(${\rm erg~s^{-1}}$)}  & \colhead{}  &  \colhead{(${\rm erg~s^{-1}}$)} &  \colhead{(${\rm erg~s^{-1}}$)}&  \colhead{}&  \colhead{}
}
\colnumbers
\startdata
J0001.2+4741	&	bcu	&	nan	&	45.029	$\pm$	0.762	&	8.88	$\pm$	1.2	&	44.009	$\pm$	0.359	&	45.069	$\pm$	0.695	&	-1.06	$\pm$	0.78	&	B3 2358+474	\\
J0001.2-0747	&	bll	&	nan	&	44.472	$\pm$	0.762	&	9	$\pm$	1.2	&	44.024	$\pm$	0.359	&	44.604	$\pm$	0.568	&	-0.58	$\pm$	0.67	&	PMN J0001-0746	\\
J0001.4-0010	&	bll	&	0.46	&	44.505	$\pm$	0.762	&	6.14	$\pm$	1.2	&	43.323	$\pm$	0.207	&	44.532	$\pm$	0.713	&	-1.21	$\pm$	0.74	&	FBQS J0001-0011	\\
J0001.5+2113	&	fsrq	&	1.11	&	45.24	$\pm$	0.762	&	18.1	$\pm$	1.2	&	44.818	$\pm$	0.0687	&	45.379	$\pm$	0.552	&	-0.561	$\pm$	0.56	&	TXS 2358+209	\\
J0001.8-2153	&	bcu	&	nan	&	45.282	$\pm$	0.762	&	7.93	$\pm$	1.2	&	43.883	$\pm$	0.359	&	45.299	$\pm$	0.731	&	-1.42	$\pm$	0.81	&	PKS 2359-221	\\
\enddata
\tablecomments{Column (1): source name of 4FGL; Column (2): class of jetted AGNs in which bll is BL Lac, fsrq is FSRQ, bcu is blazar of uncertain type, nlsy1 is NLSY1, rdg is radio galaxy, css is compact steep spectrum radio source and agn is unclassified jetted AGNs; Column (3): redshift; Column (4): logarithm of jet kinetic power assuming that $P_\mathrm{kin} = P_\mathrm{cav}$, and its corresponding uncertainty; Column (5): bulk Lorentz factor and its corresponding uncertainty; Column (6): logarithm of jet radiative power and its corresponding uncertainty; Column (7): logarithm of total jet power and its corresponding uncertainty; Column (8): logarithm of jet radiative efficiency and its corresponding uncertainty; Column (9): name of associated counterpart in other bands; The second row in the header is the unit. The ``nan" denotes no data available. AGNs with directly measured bulk Lorentz factors have an uncertainty of 0.59, while those without direct measurements have an uncertainty of 1.2. This table is available in its entirety in machine-readable form. A portion is shown here 
for guidance regarding its form and content.} 
\end{deluxetable*}

\subsection{Black hole mass, Eddington ratio and Compton dominance} 

Black hole masses and accretion‑disk luminosities for 706 blazars were taken from \citet{Paliya2021}, who classified their sample into emission‑line blazars (predominantly radiatively efficient jetted AGNs) and absorption‑line blazars (mostly radiatively inefficient jetted AGNs). These 706 blazars form a subset of our sample and were the only objects used to estimate black hole spin, accretion‑disk magnetic field, and perform the related correlation analyses.

The Eddington ratio is defined as $L_{\rm disk}/L_{\rm Edd}$, where $L_{\rm disk}=L_{\rm bol}  \approx 10L_{\rm BLR} $ \citep[e.g.,][]{Netzer1990,Baldwin1978} and $ L_{\rm Edd} = 1.3 \times 10^{38}(M/M_\odot)~{\rm erg~s^{-1}} $. For jetted AGNs, a physical-origin classification that distinguishes sources by accretion rate has been proposed \citep[e.g.,][]{Ghisellini2001,Ghisellini2011,Sbarrato2012,Xiong2014,Padovani2017}. The dividing line separating radiatively efficient and inefficient AGNs is approximately $L_{\rm disk}/L_{\rm Edd}\approx5\times10^{-3}$ \citep{Sbarrato2012}. This criterion was used to classify the sample into radiatively efficient and inefficient AGNs, not by jet radiative efficiency. Black hole masses for radiatively efficient blazars were estimated via the traditional virial method from optical spectra, using broad‑line FWHM and either continuum or $H\alpha$ line luminosities. whereas for radiatively inefficient sources masses were derived from stellar velocity dispersions or host‑bulge absolute magnitudes \citep{Paliya2021,Xiong2025}. 

For broad‑line blazars, emission‑line luminosities were obtained by fitting broad and narrow components using multi‑Gaussian and single‑Gaussian models, respectively \citep{Paliya2021}. Unlike broad-line blazars, where emission-line luminosities can be used to infer BLR luminosity, this cannot be done for absorption-line systems because broad lines are not detected. \citet{Paliya2021} therefore derived a $3\sigma$ upper limit on the H$\beta$ or Mg\,\textsc{II} line luminosity by: (1) shifting the spectrum to the rest frame and subtracting the host galaxy; (2) in the relevant wavelength range, fitting a power-law continuum plus a Gaussian emission line with fixed $\mathrm{FWHM} = 4000~\mathrm{km\,s^{-1}}$ and variable line luminosity; (3) taking as the upper limit the line luminosity at which $\chi^2$ exceeds the 99.7\% ($3\sigma$) confidence level. The BLR luminosities or upper limits were derived from the emission‑line luminosities and ratios.

For radiatively inefficient (absorption‑line) blazars, the collected disk luminosities represent upper limits, and any quantities derived from them should be treated as corresponding upper or lower limits.

\begin{deluxetable*}{cccccccccl}
\tabletypesize{\scriptsize}
\tablenum{2}
\tablewidth{0pt}
\tablecaption{Black hole spin, magnetic field of accretion disk and Compton dominance in radiatively efficient jetted active galactic nuclei. \label{tab:2}} 
\tablehead{
\colhead{4FGL name} & \colhead{$z$} &\colhead{Class} &\colhead{${\rm Log}~M_{\rm BH}$}  &  \colhead{${\rm Log}~L_{\rm disk}$} &\colhead{${\rm Log}~\frac{L_{\rm disk}}{L_{\rm Edd}}$} & \colhead{$j_{\rm cav}$}  &  \colhead{$j_{\rm Will}$} &  \colhead{${\rm Log}~B$} & \colhead{${\rm Log}~A_{\rm C}$}\\
\colhead{} & \colhead{} &  \colhead{} & \colhead{($M_{\odot}$)}  & \colhead{(${\rm erg~s^{-1}}$)} &  \colhead{} &  \colhead{}&  \colhead{}&  \colhead{(Gauss)}& \colhead{}
}
\colnumbers
\startdata
J0001.5+2113	&	1.11	&	fsrq	&	7.54	$\pm$	0.07	&	44.65	$\pm$	0.02	&	-1	$\pm$	0.0728	&	1	&	0.929	&	4.792	&	1.09	$\pm$	0.0404	\\
J0004.3+4614	&	1.81	&	fsrq	&	8.36	$\pm$	0.1	&	46.07	$\pm$	0.03	&	-0.404	$\pm$	0.104	&	0.885	&	0.211	&	4.511	&	0.947	$\pm$	0.143	\\
J0004.4-4737	&	0.88	&	fsrq	&	8.28	$\pm$	0.27	&	45.1	$\pm$	0.1	&	-1.29	$\pm$	0.288	&	1	&	0.474	&	4.36	&	-0.144	$\pm$	0.184	\\
J0006.3-0620	&	0.35	&	bll	&	8.93	$\pm$	0.4	&	44.52	$\pm$	0.15	&	-2.52	$\pm$	0.427	&	0.971	&	0.258	&	3.771	&	-1.5	$\pm$	0.119	\\
J0010.6+2043	&	0.6	&	fsrq	&	7.86	$\pm$	0.04	&	45.35	$\pm$	0.03	&	-0.624	$\pm$	0.05	&	0.963	&	0.22	&	4.714	&	0.116	$\pm$	0.411	\\
\enddata
\tablecomments{Column (1): source name of 4FGL; Column (2): class of jetted AGNs in which bll is BL Lac, fsrq is FSRQ, bcu is blazar of uncertain type, nlsy1 is NLSY1, rdg is radio galaxy, css is compact steep spectrum radio source and agn is AGN; Column (3): redshift; Column (4): logarithm of black hole mass and its corresponding uncertainty; Column (5): logarithm of accretion disk luminosity and its corresponding uncertainty; Column (6): logarithm of Eddington ratio and its corresponding uncertainty; Column (7): dimensionless black hole spin assuming that $P_\mathrm{kin} = P_\mathrm{cav}$; Column (8): dimensionless black hole spin assuming that $P_\mathrm{kin} = P_\mathrm{W}$ with $f=1$; Column (9): logarithm of magnetic field of accretion disk; Column (10): logarithm of Compton dominance and its corresponding uncertainty. The ``nan" denotes no data available. The black hole masses were estimated using the single-epoch virial method, as reported by \citet{Paliya2021}, \citet{Foschini2015}, \citet{Paliya2024} and \citet{Zhang2020}. This table is available in its entirety in machine-readable form. A portion is shown here 
for guidance regarding its form and content.} 
\end{deluxetable*}

\begin{deluxetable*}{cccccccccl}
\tabletypesize{\scriptsize}
\tablenum{3}
\tablewidth{0pt}
\tablecaption{Black hole spin, magnetic field of accretion disk and Compton dominance in radiatively inefficient jetted active galactic nuclei. \label{tab:3}} 
\tablehead{
\colhead{4FGL name} & \colhead{$z$} &\colhead{Class} &\colhead{${\rm Log}~M_{\rm BH}$}  &  \colhead{${\rm Log}~L_{\rm disk}$ (UL)} &\colhead{${\rm Log}~\frac{L_{\rm disk}}{L_{\rm Edd}}$ (UL)} & \colhead{$j_{\rm cav}$ (LL)}  &  \colhead{$j_{\rm Will}$ (LL)} &  \colhead{${\rm Log}~B$ (UL)}&\colhead{${\rm Log}~A_{\rm C}$}\\
\colhead{} & \colhead{} &  \colhead{} & \colhead{($M_{\odot}$)}  & \colhead{(${\rm erg~s^{-1}}$)} &  \colhead{} &  \colhead{}&  \colhead{}&  \colhead{(Gauss)}& \colhead{}
}
\colnumbers
\startdata
J0009.7-3217	&	0.03	&	rdg	&	8.811	$\pm$	nan	&	41.709	&	-5.22	&	0.94	&	0.11	&	3.251	&	-0.637	$\pm$	0.144	\\
J0013.9-1854	&	0.09	&	bll	&	9.65	$\pm$	0.19	&	43.27	&	-4.49	&	0.243	&	0.024	&	2.875	&	-1	$\pm$	0.122	\\
J0014.2+0854	&	0.16	&	bll	&	8.85	$\pm$	0.12	&	43.37	&	-3.59	&	0.977	&	0.256	&	3.491	&	nan	$\pm$	nan	\\
J0017.8+1455	&	0.3	&	bll	&	8.27	$\pm$	0.44	&	44.16	&	-2.22	&	0.704	&	0.0944	&	4.109	&	-0.435	$\pm$	0.101	\\
J0021.6-0855	&	0.65	&	bll	&	8.54	$\pm$	0.45	&	44.63	&	-2.02	&	0.916	&	0.209	&	4.022	&	nan	$\pm$	nan	\\
\enddata
\tablecomments{The information on this table is the same as in the Table 2 except that Columns (5)–(9) list upper limit (UL) or lower limit (LL). The black hole masses were derived from stellar velocity dispersions or host-bulge magnitudes, as reported by \citet{Paliya2021}, \citet{Chen2023}. This table is available in its entirety in machine-readable form. A portion is shown here 
for guidance regarding its form and content.} 
\end{deluxetable*}

Black hole masses for 12 radio galaxies were acquired through cross-matching with the gamma‑ray–emitting radio galaxy sample of \citet{Chen2023} and were estimated from the host bulge near‑infrared magnitudes. \citet{Sbarrato2014} investigated the correlation between accretion rate and jet power for samples of blazars and radio galaxies, deriving the following relationship:
\begin{equation}
    \log \frac{L_{\text{BLR}}}{L_{\text{Edd}}} = 0.87 (\pm 0.12) \log \frac{L_{\text{8GHz}}}{L_{\text{Edd}}} + 0.28.
    \label{eq:placeholder_label}
\end{equation}
On one hand, the radio galaxies in the sample are radiatively inefficient, with $\frac{L_{\text{BLR}}}{L_{\text{Edd}}}$ representing an upper limit. On the other hand, a broken power-law with steeper slope better describes the data (see Fig.3 in \citet{Sbarrato2014}). Therefore, the estimated disk luminosity and Eddington ratio obtained from this relationship should be considered as upper limits. Radio data for 11 radio galaxies were sourced from the Combined Radio All-Sky Targeted Eight GHz Survey (CRATES; \cite{Healey2007}). For 4 sources not included in CRATES, measurements at 8 GHz or 5 GHz were retrieved from NASA/NED. The 5 GHz radio fluxes were converted to 8 GHz assuming a spectral index of $\alpha=0.0$. For the 3 radio galaxies lacking black hole mass measurements, the sample mean black hole mass for radio galaxies was used, and only radio galaxies with measured black hole masses were included in correlation analyses.

\begin{figure*}
\centering
\gridline{\fig{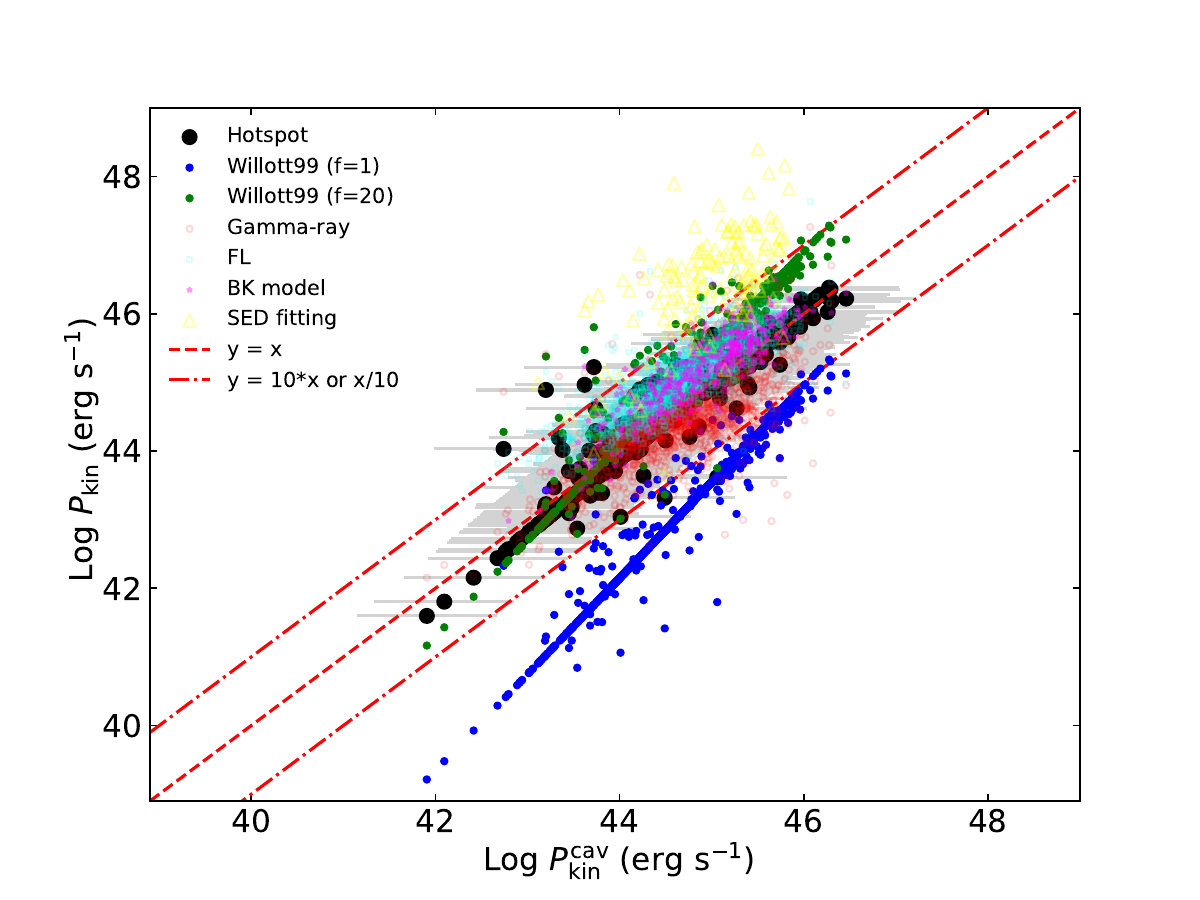}{0.48\textwidth}{(a) Jet cavity power versus jet kinetic powers estimated by other methods. The horizontal axis displays jet cavity power, while the vertical axis shows jet powers estimated by other methods (color coded: Hotspot—method of hotspots, Willott99—\citet{Willott1999}, Gamma-ray—from gamma ray, FL—fundamental line/plane, BK model—Blandford and K\"onigl, SED fitting—SED‑derived). The gray short line indicates the uncertainty in the cavity power. The red dashed line marks the $y = x$ locus, and the dash‑dot lines mark $y = 10x$ and $x/10$.} \label{fig:3a}
          \fig{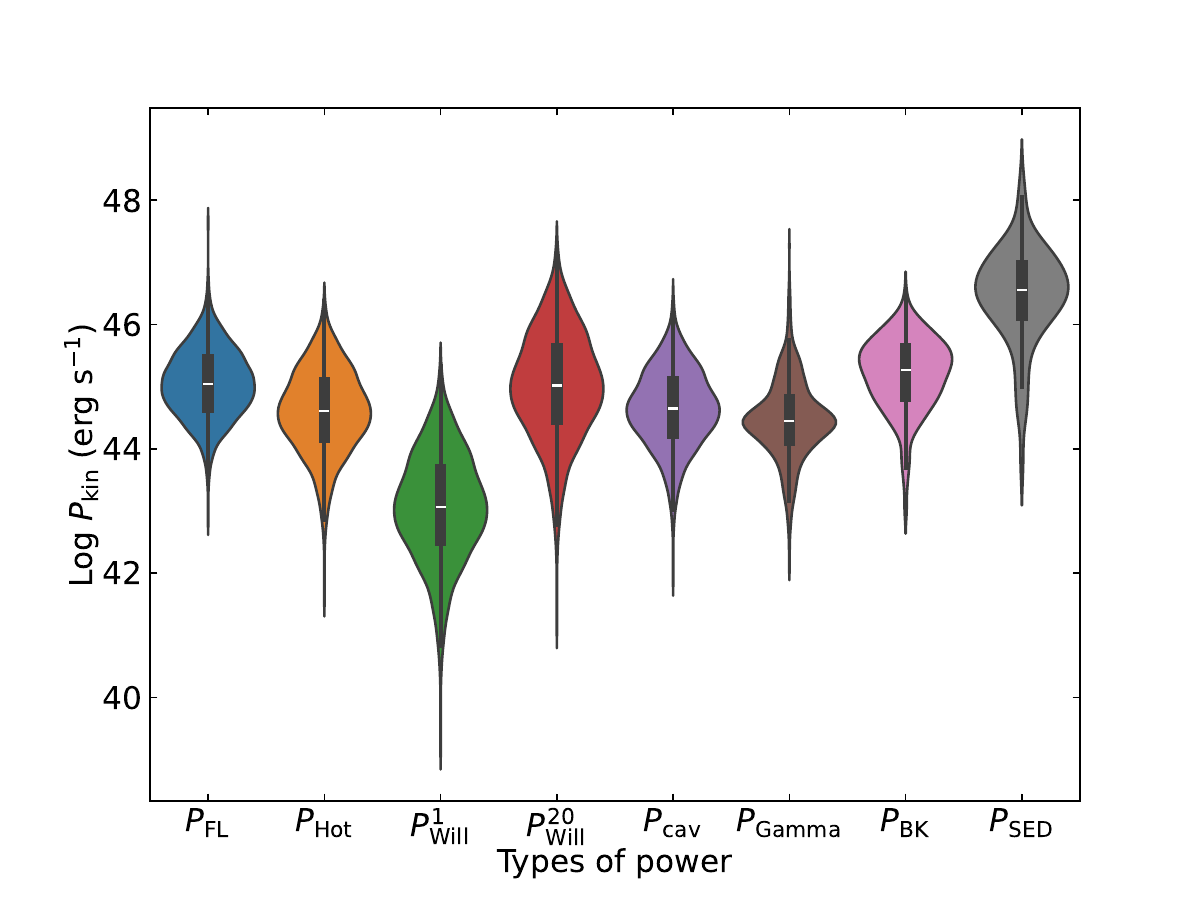}{0.48\textwidth}{(b) Violin plot of jet power distributions. The different colors indicate jet kinetic powers derived using different methods. The central line and box indicate the median and quartiles, respectively. Symmetrical, violin-shaped outlines display the kernel density estimate of the data.}\label{fig:3b}}
    \caption{Comparative analysis of jet kinetic powers in blazars.}
    \label{fig:Comparative analysis}
\end{figure*} 

Black hole masses and accretion‑disk luminosities for the 4 NLSy1 were taken from \citet{Foschini2015}, while the remaining 2 were from \citet{Paliya2024}. Both studies employed the single‑epoch virial technique to estimate black hole masses. The accretion‑disk luminosities were determined from the $H{\rm \beta}$ line luminosity and from the 5100~\AA\ continuum luminosity, respectively.

Black hole masses and disk luminosities for the 5 CSS sources were adopted from \citet{Zhang2020}; the masses were taken from \citet{Gu2001,Shen2011,Woo2002} and were estimated via the single‑epoch virial method, and disk luminosities were estimated from the narrow blue bump in the SEDs or from BLR luminosities following \citet{Celotti1997}. Black hole masses and accretion-disk bolometric luminosities of the 2 SSRQ were derived from \citet{Paliya2024} and \citet{Li2021}, respectively.

Black hole masses and disk luminosities for the 4 unclassified jetted AGNs are unavailable in the literature; therefore, these sources will be excluded from analyses involving parameters associated with black hole mass and disk luminosity.

Black hole masses are determined with different methods (some sources have spectral lines and some don't), so this might cause some biases to correlations.

The Compton dominance can be expressed as
\begin{equation}
    A_C = \frac{L^{\text{obs}}_{\text{IC}}}{L^{\text{obs}}_{\text{syn}}}=\frac{\nu F_{\nu}~{\rm (inverse~Compton~peak)}}{\nu F_{\nu}~{\rm (synchrotron~peak)}},
\end{equation}
following the formulation in \citet{Finke2013}. The $\nu F_{\nu}$ values at synchrotron-peak frequency and high-energy inverse Compton peak frequency were extracted from Table A1 of 4LAC-DR3. 

\subsection{Black hole spin and magnetic field of accretion disk} \label{subsec:spin}

The “outflow method" was applied to estimate the black hole spin. When this method was initially proposed, it was applicable only to FR II radio sources owing to the constraints of the strong shock approach \citep{Daly2016,Daly2019}. After accounting for the fundamental line/plane of black hole \citep{Daly2018} to estimate jet kinetic power (or beam power), \citet{Daly2019} also inferred the black hole spin for stellar-mass galactic black holes, compact radio sources, and low-ionization nuclear emission-line regions (LINERs), in addition to FR II radio sources. The black hole fundamental plane (FP) was first proposed and thoroughly analyzed by \citet{Merloni2003} and \citet{Falcke2004}. Since then, numerous studies have expanded the FP using larger and more diverse samples, including \citet{Kording2006}, \citet{Gultekin2009}, \citet{Bonchi2013}, \citet{Saikia2015}, and \citet{Nisbet2016}. The FP encapsulates the relationship between radio luminosity, X-ray luminosity, and black hole mass for black hole systems and provides a good description of the data over a very large range of black hole mass \citep{Daly2018}. A caveat regarding the FP is that both radio and X‑ray luminosities depend on redshift.

Our sample concentrates on gamma-ray–emitting jetted AGNs, characterized by powerful relativistic jets. Therefore, following the estimation of jet cavity power, the “outflow method” can be employed to infer the spin of the black hole.

The key assumption of the “outflow method" is that the jet or outflow is at least partially powered by the black hole's spin, as described by \cite{Blandford1977,Meier1999}. In one representation of the generalized Blandford-Znajek (BZ) model, the spin function $f(j)$ is converted to the dimensionless black hole spin $j$ via $\sqrt{\frac{f(j)}{f_{\text{max}}}} = j \left(1 + \sqrt{1 - j^2}\right)^{-1}$, where $j\equiv Jc/(GM^2)$, with $J$ being the black hole’s spin angular momentum, $M$ its mass, $c$ the speed of light, and $G$ Newton’s constant. Thus, for $f(j) / f_{\text{max}}\leq 1$, the spin can be expressed as 
\begin{equation}
    j = \frac{2 \sqrt{f(j) / f_{\text{max}}}}{f(j) / f_{\text{max}} + 1},
\end{equation}
where $f_{\text{max}}$ is the value of $f(j)$ when $j=1$. Any $j$ values for which $f(j) / f_{\text{max}} > 1$ are capped at 1 \citep{Daly2019}. The normalized spin function is written as
\begin{equation}
    \frac{f(j)}{f_{\text{max}}} = \left( \frac{L_j}{g_j L_{\text{Edd}}} \right) \left( \frac{L_{\text{disk}}}{g_{\text{bol}} L_{\text{Edd}}} \right)^{-A}.
\end{equation}
Here, $L_j$ is the jet kinetic power. Normalization factors $g_{\rm bol} = 1$ and $g_{\rm j} = 0.1$ were adopted, as suggested by \citet{Daly2019,Daly2016,Daly2018}. \citet{Daly2016,Daly2019} derived an empirical relation linking jet kinetic power, accretion‑disk luminosity $L_{\rm disk}$, and Eddington luminosity $L_{\rm Edd}$ of the form
\begin{equation}
    \frac{L_j}{L_{\text{disk}}} \propto \left( \frac{L_{\text{disk}}}{L_{\text{Edd}}} \right)^{A-1}.
\end{equation}
Combining this with equation (5) of \citet{Fan2019} yields $A = 0.43\pm0.02$. This spin estimate uses as inputs the jet cavity power from extended 300 MHz flux ($P_\mathrm{kin} = P_\mathrm{cav}$), accretion‑disk luminosity, and black hole mass.

The total magnetic field strength $B$ of accretion disk was obtained from the relation \citep{Daly2019}:
\begin{equation}
\begin{aligned}
    \left( \frac{B}{10^4 \, \text{G}} \right) = \left( \frac{B}{B_{\text{Edd}}} \right) \left( \frac{\kappa_B^2}{M_8} \right)^{1/2}\\ = \left( \frac{L_{\text{disk}}}{g_{\text{bol}} L_{\text{Edd}}} \right)^{A/2} \left( \frac{\kappa_B^2}{M_8} \right)^{1/2},
\end{aligned}
\end{equation}
where the Eddington magnetic field strength in units of $10^4$ G is given by $B_{\text{Edd,4}} = \kappa_{\rm B} M_{\rm 8}^{-1/2}$, with $\kappa_{\rm B}\simeq 6$ and black hole mass $M$ in units of $10^8 M_\odot$. Estimating the accretion‑disk magnetic field requires the black hole mass and the accretion‑disk luminosity as inputs.

The key parameters for gamma-ray emitting jetted AGNs are summarized in Tables 1-3.

\section{Results} \label{sec:Results}

\subsection{Comparative analysis of jet kinetic powers in blazars}

Figure 3 presents the results of the comparative analysis between the jet cavity power and the jet kinetic power estimated by other methods. As shown in Fig. 3, except for the jet power obtained through the SED fitting method and Willott et al.'s approach ($f=1$), the jet cavity power for most of the sample is consistent with the estimates from other methods (i.e., Willott et al.'s model-dependent approach with $f=20$, hotspots method, gamma-ray luminosity estimation, Blandford and K\"onigl model, fundamental line/plane of black hole) within the measurement uncertainties. 

The jet power estimated via the hotspots method is nearly identical to the cavity power with the exception of a few data points. Additionally, the jet power estimated with Willott et al.'s approach ($f = 1$) is among the lowest of all methods. For the majority of sample, the jet power estimated via the SED fitting method exceeds that obtained by other methods by one or more ordes of magnitude, even when accounting for uncertainty. Based on the best-fit linear relation, the correlation between jet cavity power and kinetic power estimated via SED fitting can be expressed as 
\begin{equation}
    \log P_{\rm SED} = 0.876(\pm 0.09) \log P_{\rm cav} + 7.06(\pm 4.09)
\end{equation}
with scatter of 0.37 dex.

\begin{figure}
    \centering
    \includegraphics[width=0.5\textwidth]{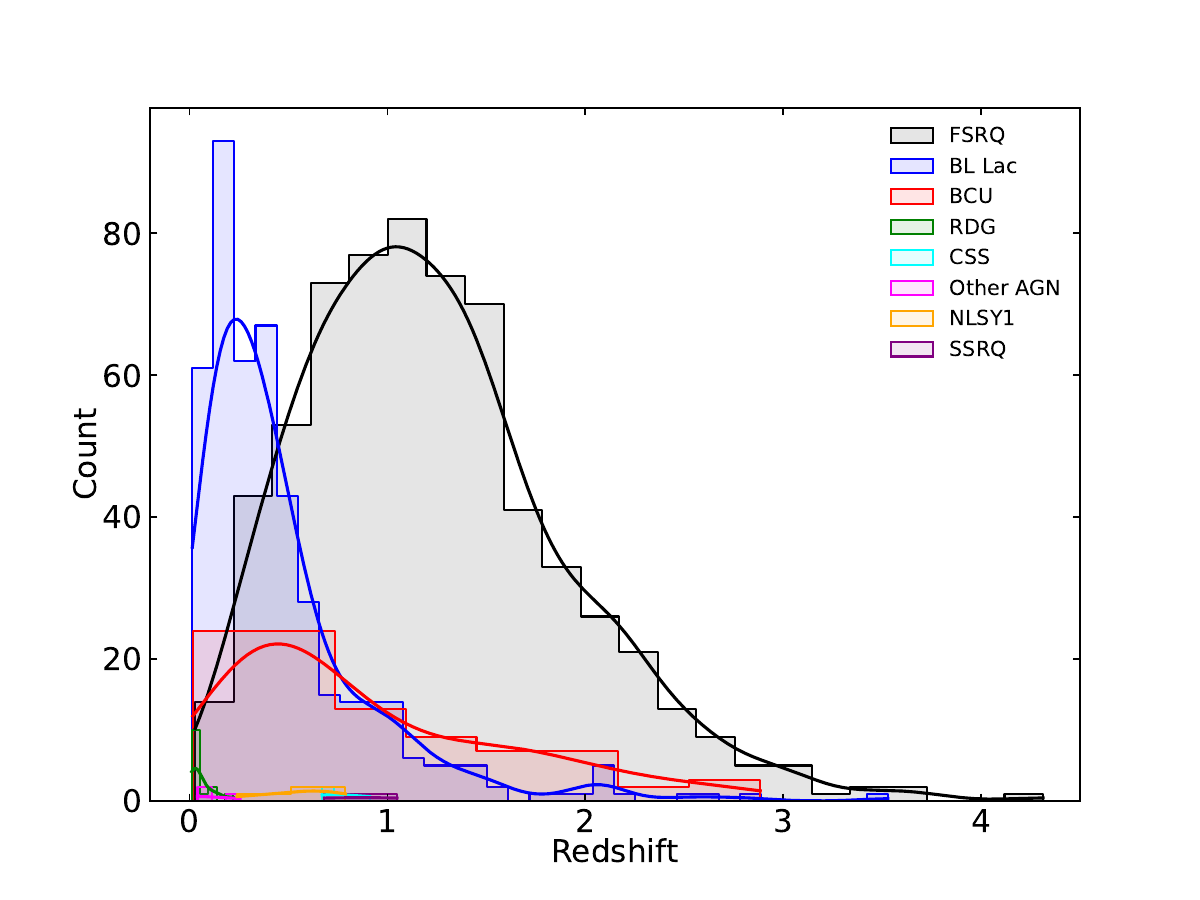}
    \caption{The distributions of redshift for jetted AGNs. The histograms, filled with different colors, represent the redshift distributions for various categories of jetted AGNs. Kernel density estimates are overlaid to smooth these distributions, represented by lines of corresponding colors.}
\end{figure}

\subsection{The distributions of the parameters}

The redshift distribution for FSRQs spans from 0.028 to 4.3, with a mean value of 1.2. In comparison, BL Lac objects range from 0.01 to 3.5, with an average redshift of 0.47. The mean redshift for BCUs is 0.87. The redshift distribution for other types of jetted AGNs is generally below 1.1, as illustrated in Fig. 4. 

\begin{figure*}
\centering
\gridline{\fig{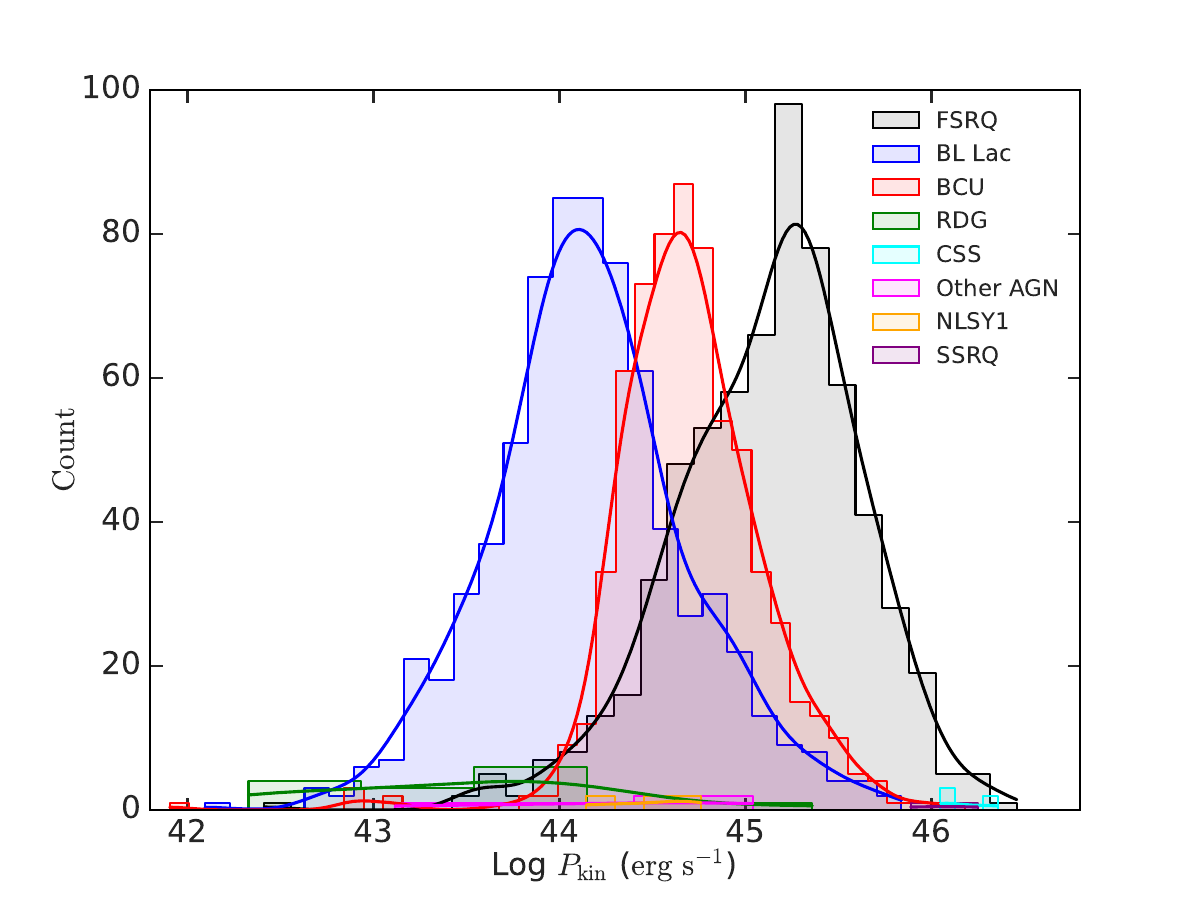}{0.52\textwidth}{(a) Jet kinetic power} 
          \fig{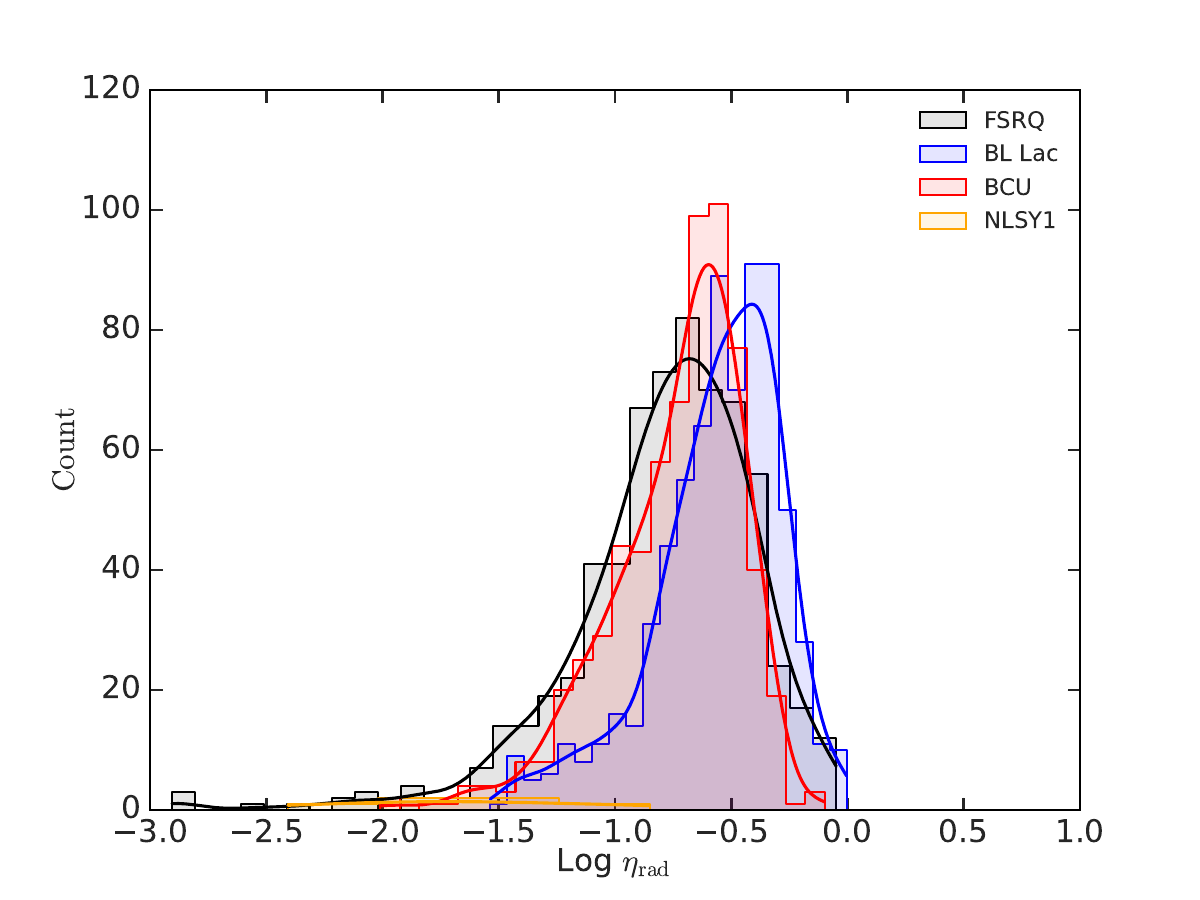}{0.52\textwidth}{(b) Jet radiative efficiency}}
    \caption{The distributions of jet kinetic power and radiative efficiency. The other designations are the same as those in Fig. 4.}
\end{figure*} 

The distributions of jet kinetic powers are shown in Fig. 5a. Unless otherwise noted, jet kinetic power refers to the jet cavity power. The jet kinetic powers span $10^{42.41}-10^{46.46}~{\rm erg~s^{-1}}$ for FSRQ (with an average of $10^{45.12 \pm0.02}~{\rm erg~s^{-1}}$), $10^{42.09}–10^{45.84}~{\rm erg~s^{-1}}$ for BL Lac (average of $10^{44.16 \pm0.02}~{\rm erg~s^{-1}}$), $10^{41.91}–10^{46.18}~{\rm erg~s^{-1}}$ for BCU (average of $10^{44.70 \pm0.02}~{\rm erg~s^{-1}}$), $10^{42.32}–10^{45.36}~{\rm erg~s^{-1}}$ for RDG (average of $10^{43.52 \pm0.21}~{\rm erg~s^{-1}}$), $10^{46.05}–10^{46.36}~{\rm erg~s^{-1}}$ for CSS (average of $10^{46.17 \pm0.06}~{\rm erg~s^{-1}}$), $10^{43.12}–10^{45.04}~{\rm erg~s^{-1}}$ for unclassified jetted AGNs (average of $10^{44.24 \pm0.37}~{\rm erg~s^{-1}}$), $10^{44.15}–10^{44.76}~{\rm erg~s^{-1}}$ for NLSY1 (average of $10^{44.49 \pm0.1}~{\rm erg~s^{-1}}$), and $10^{45.89}–10^{46.25}~{\rm erg~s^{-1}}$ for SSRQ (average of $10^{46.07 \pm0.13}~{\rm erg~s^{-1}}$), respectively. Therefore, the mean jet kinetic powers for gamma‑ray–emitting AGNs rank as: ${\rm CSS>SSRQ>FSRQ>BCU>NLSY1>Other~AGN}$ ${\rm >BL~Lac>RDG}$; the jet kinetic powers of both blazars and misaligned AGNs lie in the range $10^{41.91}–10^{46.46}~{\rm erg~s^{-1}}$, with overlapping regions between the populations.

The lower-limit jet radiative efficiency, computed via equation (12), has a mean of $0.033\pm0.001$, whereas the mean radiative efficiency is $0.205\pm 0.004$ (Fig. 5b), indicating that jets in gamma‑ray–emitting AGNs are dominated by kinetic rather than radiative processes. The mean jet radiative efficiencies rank as: ${\rm BL~Lac>BCU}$ ${\rm FSRQ>NLSY1}$. We cannot draw firm conclusions for the other AGN classes due to the large uncertainties in their radiative power estimates.

\begin{figure*}
\centering
\gridline{\fig{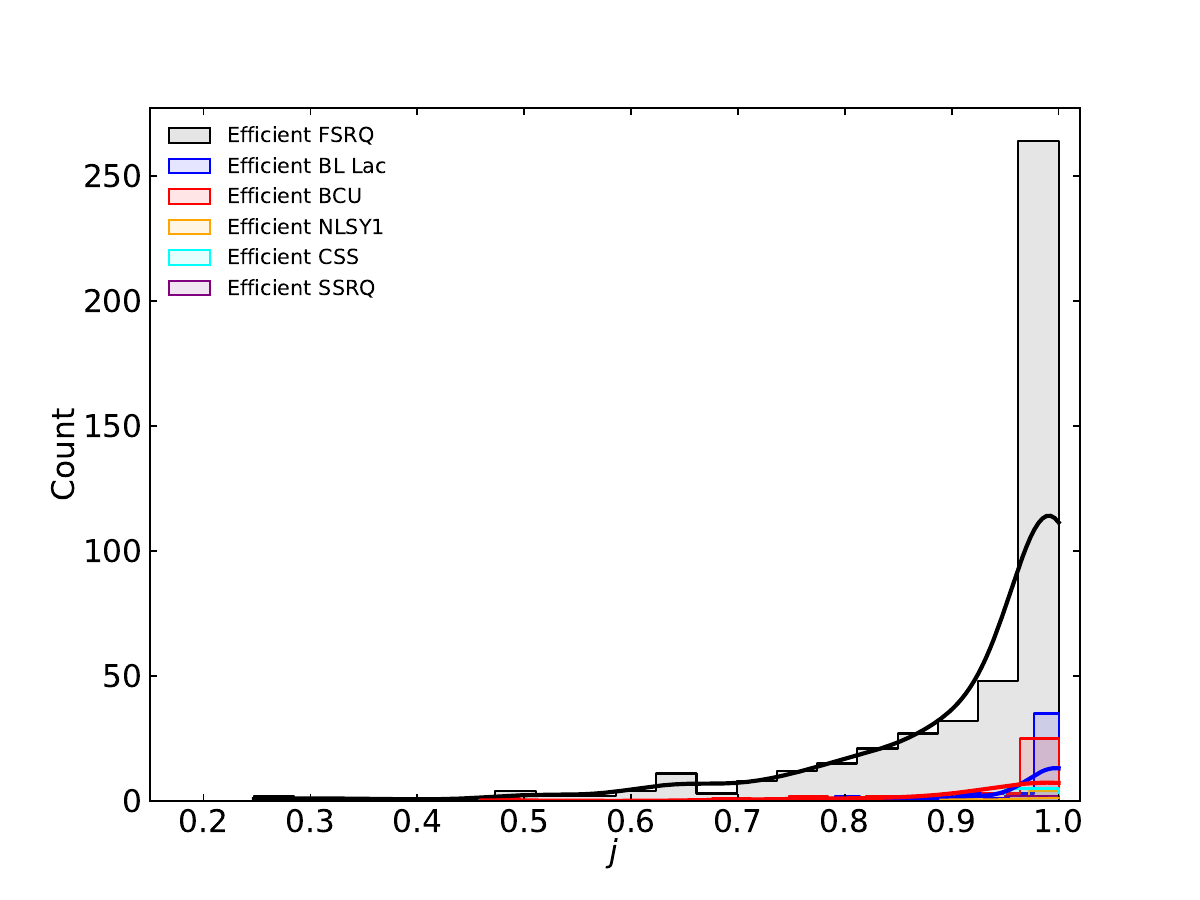}{0.52\textwidth}{(a) Radiatively efficient AGNs} 
          \fig{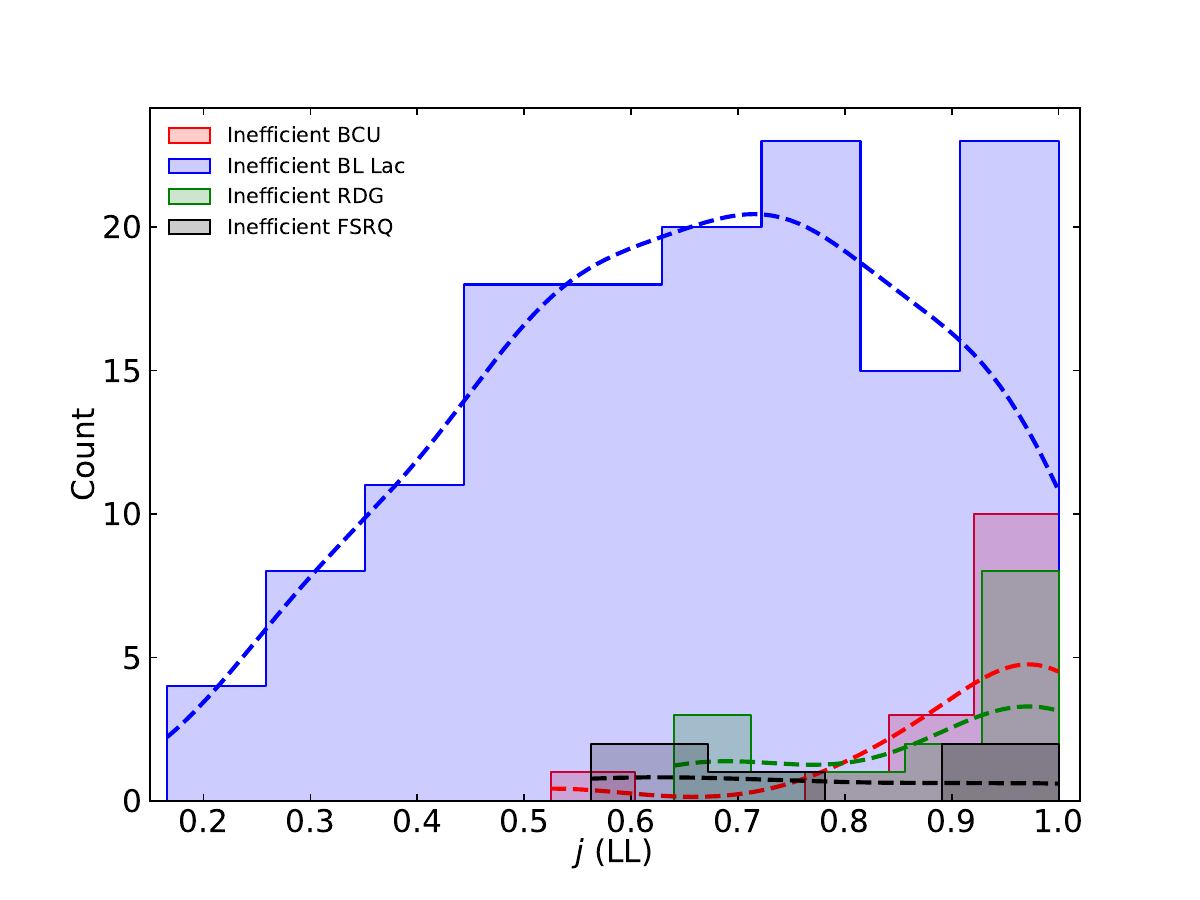}{0.52\textwidth}{(b) Radiatively inefficient AGNs}}
    \caption{The distributions of black hole spins. For radiatively inefficient AGNs, the inferred spin $j$ should be treated as a lower limit. The other designations are similar to those in Fig. 4.}
\end{figure*} 

The radiatively efficient and inefficient populations contain 554 and 175 AGNs, respectively. For radiatively efficient AGNs, the majority of sources exhibit black hole spin values exceeding 0.9, with an average spin of $0.929 \pm 0.005$. In contrast, for radiatively inefficient AGNs, the average lower limit of the black hole spin is 0.71; it is highly probable that many of these sources have high spins ($>0.9$), though some may exhibit different spin states. Thus, black hole spin does not effectively distinguish between different jetted AGN subclasses (see Fig. 6). A clear caveat: estimating black hole spin via the outflow method requires jet kinetic power, black hole mass, and accretion‑disk luminosity, and uncertainties in these inputs propagate into the spin uncertainty (see also Section 4.3).

\begin{figure*}
\centering
\gridline{\fig{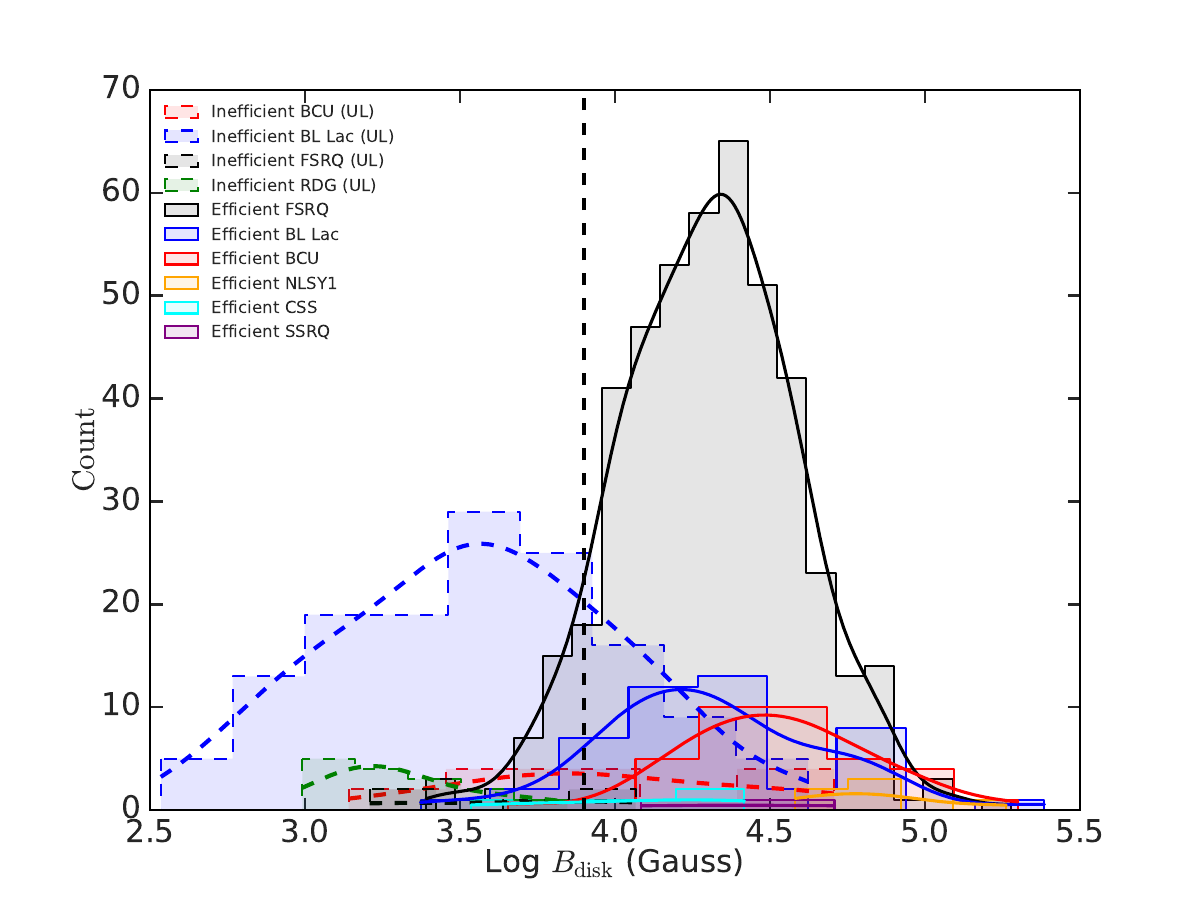}{0.52\textwidth}{(a) Total magnetic field strength of accretion disk. For radiatively inefficient AGNs, the total magnetic field strength of accretion disk is an upper limit. The solid line indicates radiatively efficient AGNs, while the dashed line represents radiatively inefficient AGNs. The vertical line indicates a magnetic field strength of $10^{3.9}$ Gauss.} 
          \fig{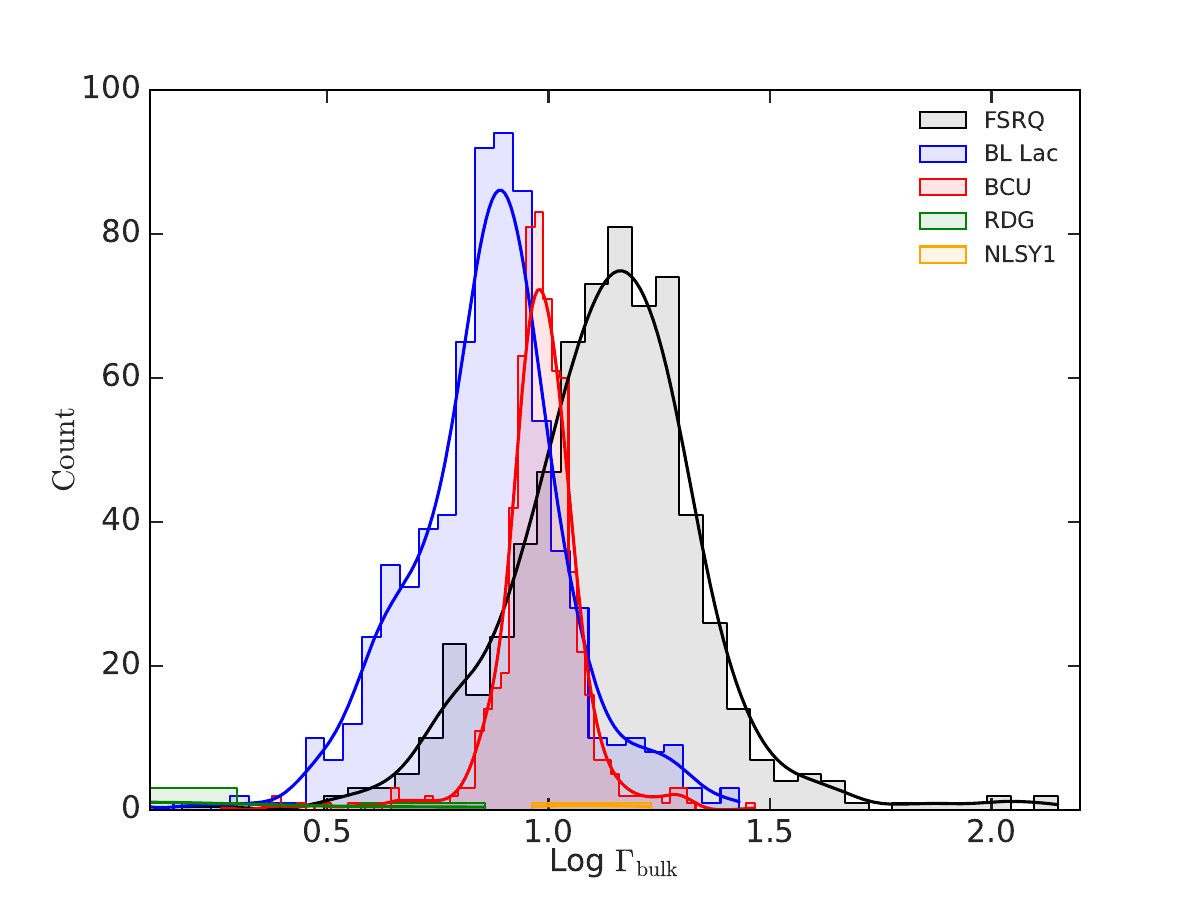}{0.5\textwidth}{(b) Bulk Lorentz factor. For non‑blazar AGNs, the bulk Lorentz factor is derived solely from the radio variability method.}}
    \caption{The distributions of total magnetic field strength of accretion disk and bulk Lorentz factor. The other designations are similar to those in Fig. 4.}
\end{figure*} 

\begin{figure*}
\centering
\gridline{\fig{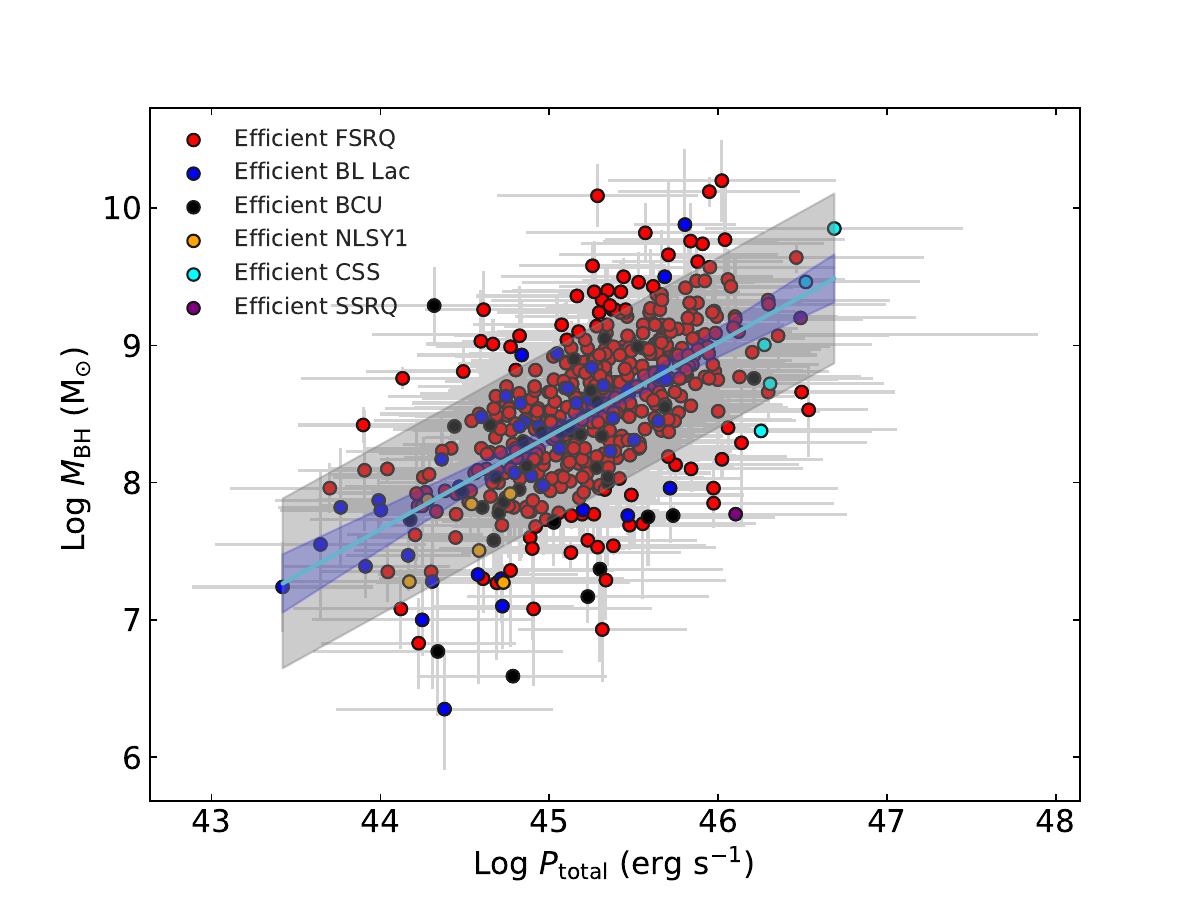}{0.52\textwidth}{(a) Black hole mass versus total jet power in radiatively efficient AGNs} 
          \fig{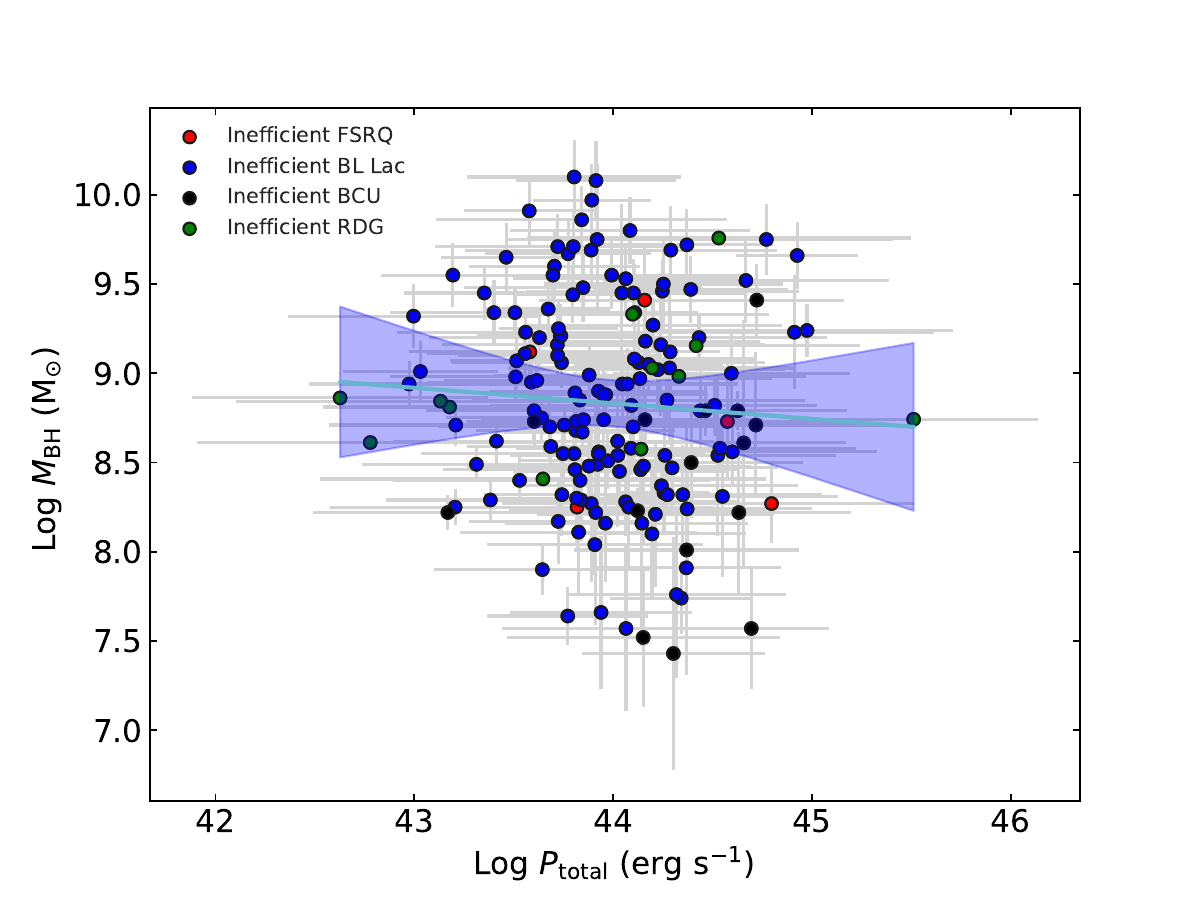}{0.52\textwidth}{(b) Black hole mass versus total jet power in radiatively inefficient AGNs}}
\gridline{\fig{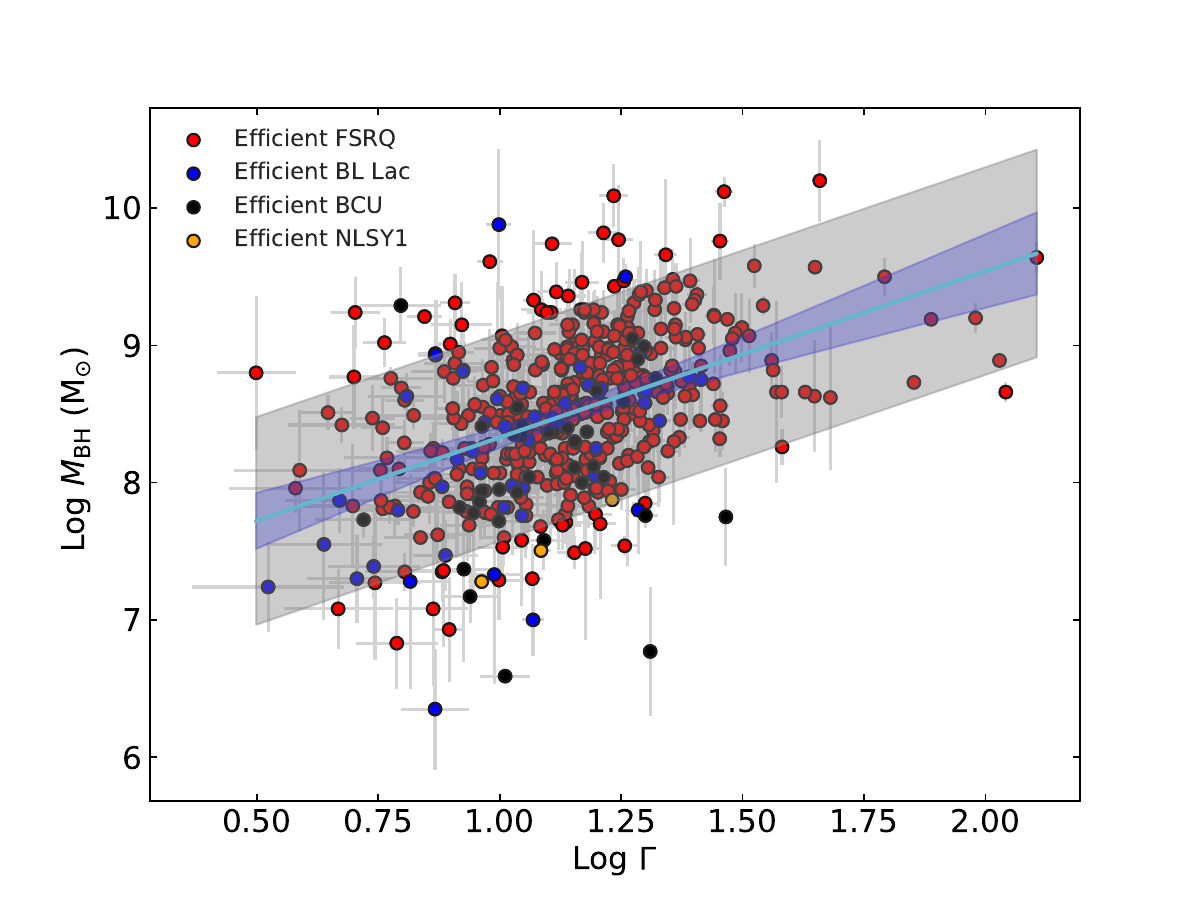}{0.52\textwidth}{(c) Black hole mass versus bulk Lorentz factor in radiatively efficient AGNs}
          \fig{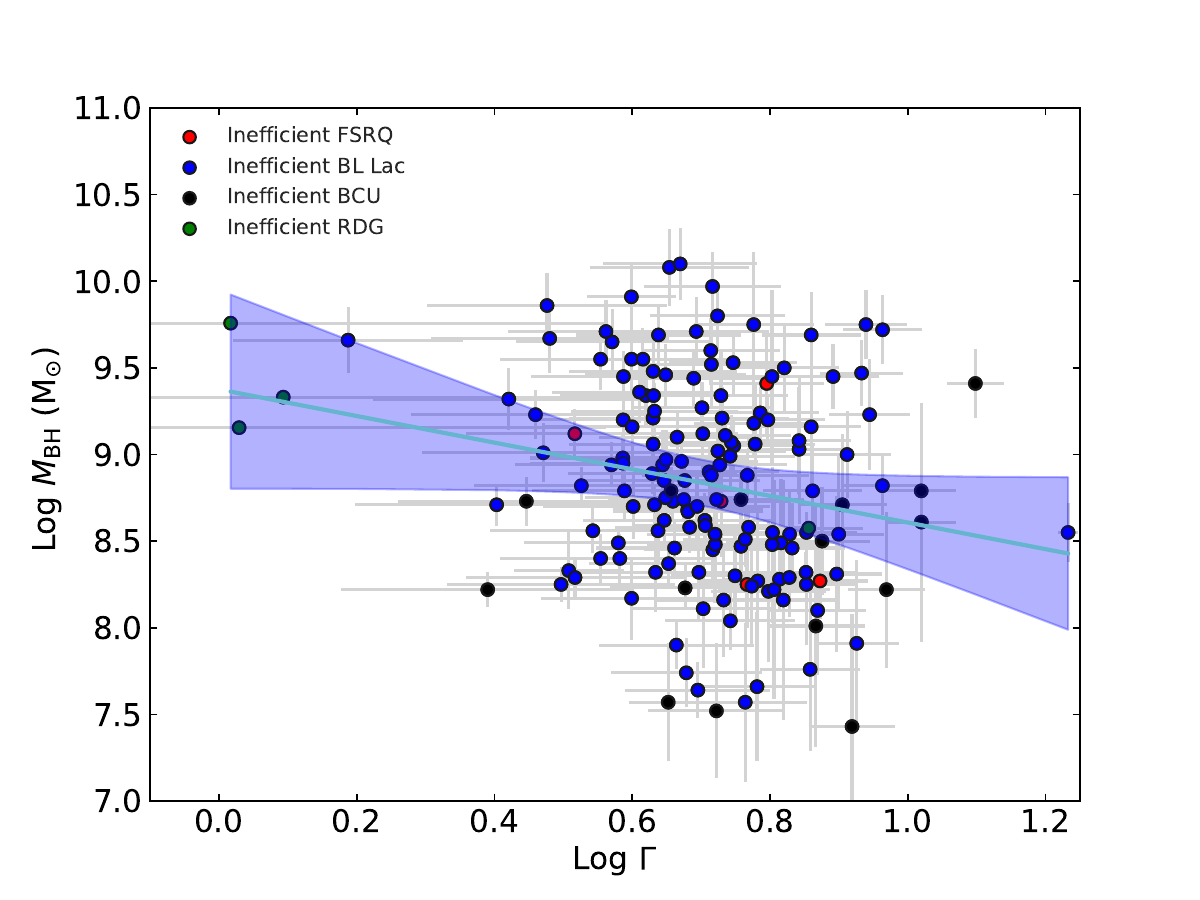}{0.52\textwidth}{(d) Black hole mass versus bulk Lorentz factor in radiatively inefficient AGNs}}
\caption{Black hole mass as a function of total jet power and bulk Lorentz factor. The blue solid line represents the best-fit linear model obtained through ordinary least squares fitting. The blue shaded region indicates the 3$\sigma$ confidence band, while the gray region depicts the 3$\sigma$ dispersion. Data points in different colors represent different subclasses of jetted AGNs.}
\end{figure*}

Fig. 7 shows the distributions of total magnetic field strength of accretion disk and bulk Lorentz factor. In radiatively efficient AGNs, the total magnetic field strength of the accretion disk spans $10^{3.3}-10^{5.5}$ Gauss, with a mean of $10^{4.32\pm0.01}$ Gauss, whereas in radiatively inefficient AGNs, the upper limit of the magnetic field strength averages $10^{3.57}$ Gauss.
Considering measurement uncertainties and the fact that ineffective AGNs represent upper limits, the dividing line between radiatively efficient and inefficient AGNs based on magnetic field strength can be approximated as $10^{3.9}$ Gauss. Radiatively efficient and inefficient AGNs still overlap near the dividing line, indicating a smooth rather than sharp transition. The magnetic field strength of the accretion disk in most radiatively efficient AGNs exceeds that in radiatively inefficient AGNs, as shown in Fig. 7a. Estimating the accretion‑disk magnetic field requires the black hole mass and the Eddington ratio (or accretion‑disk luminosity) as inputs. Black hole masses do not differ significantly between the two groups, suggesting that differing Eddington ratios drive the disparity. For a given black hole mass, accretion amplifies the disk magnetic field; higher Eddington ratios correspond to stronger disk fields.

The mean bulk Lorentz factors follow the order: ${\rm FSRQ>NLSY1>BCU>BL~Lac>RDG}$ (Fig. 7b), with FSRQ and BL Lac means of $15.7\pm0.5$ and $7.9\pm0.1$, respectively. A significant caveat: the RDG and NLSy1 samples with measured bulk Lorentz factors are very small.

\begin{figure*}
\plotone{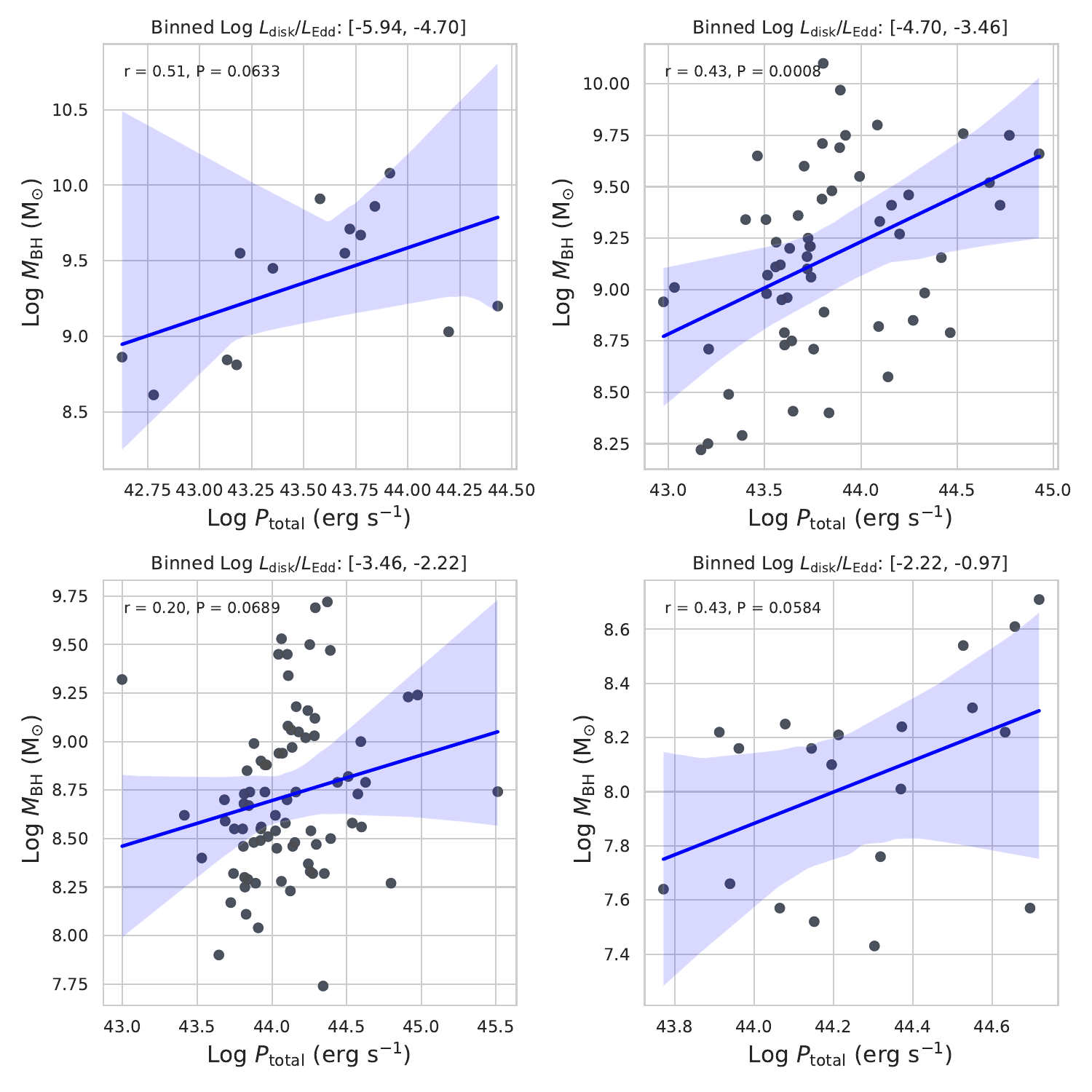}
\caption{Black hole mass versus total jet power across accretion-rate bins for radiatively inefficient AGNs. The lines and shaded areas have the same meaning as in Fig. 8. The $r$ and $P$ values in the upper‑left corner denote the correlation coefficient and its significance level, respectively.
}
\end{figure*}

\begin{figure*}
\centering
\gridline{\fig{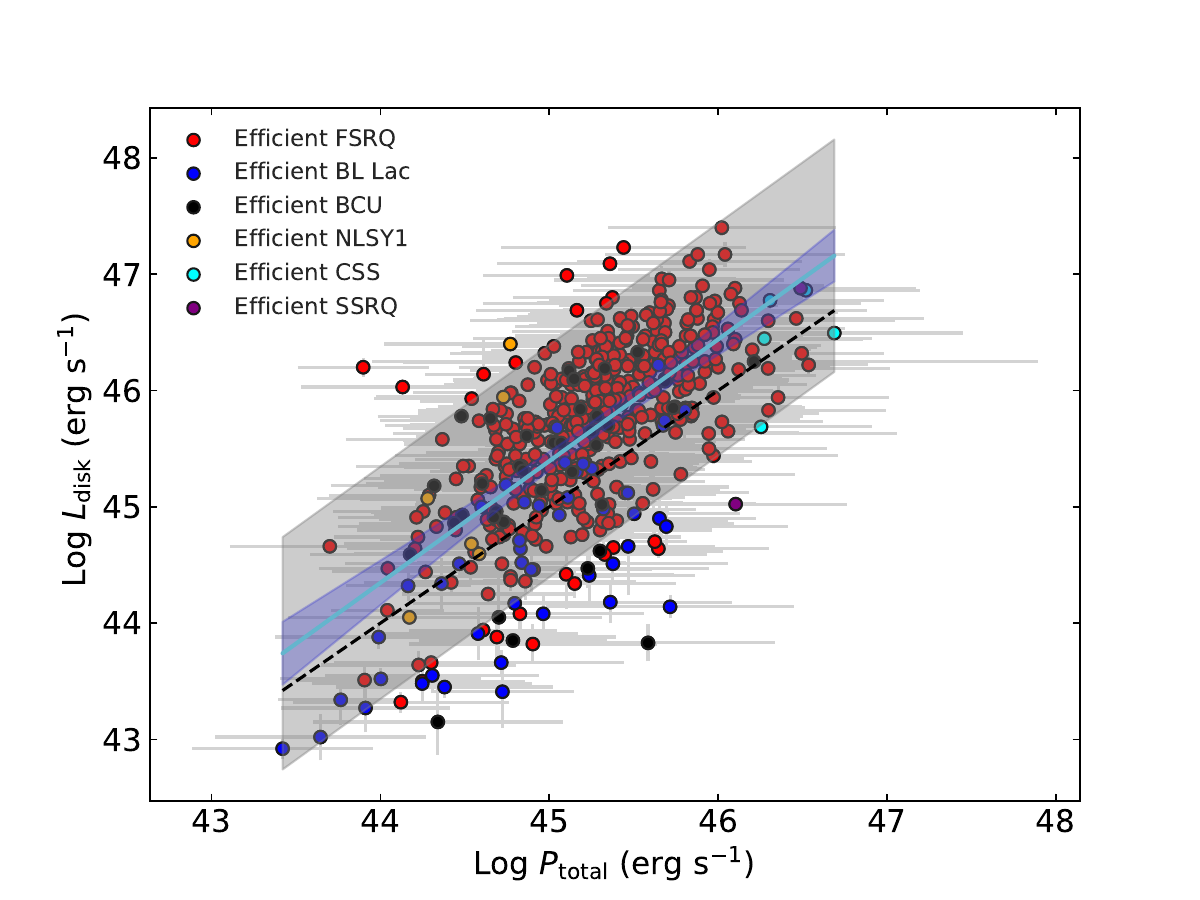}{0.52\textwidth}{(a) Total jet power versus accretion disk luminosity in radiatively efficient AGNs} 
          \fig{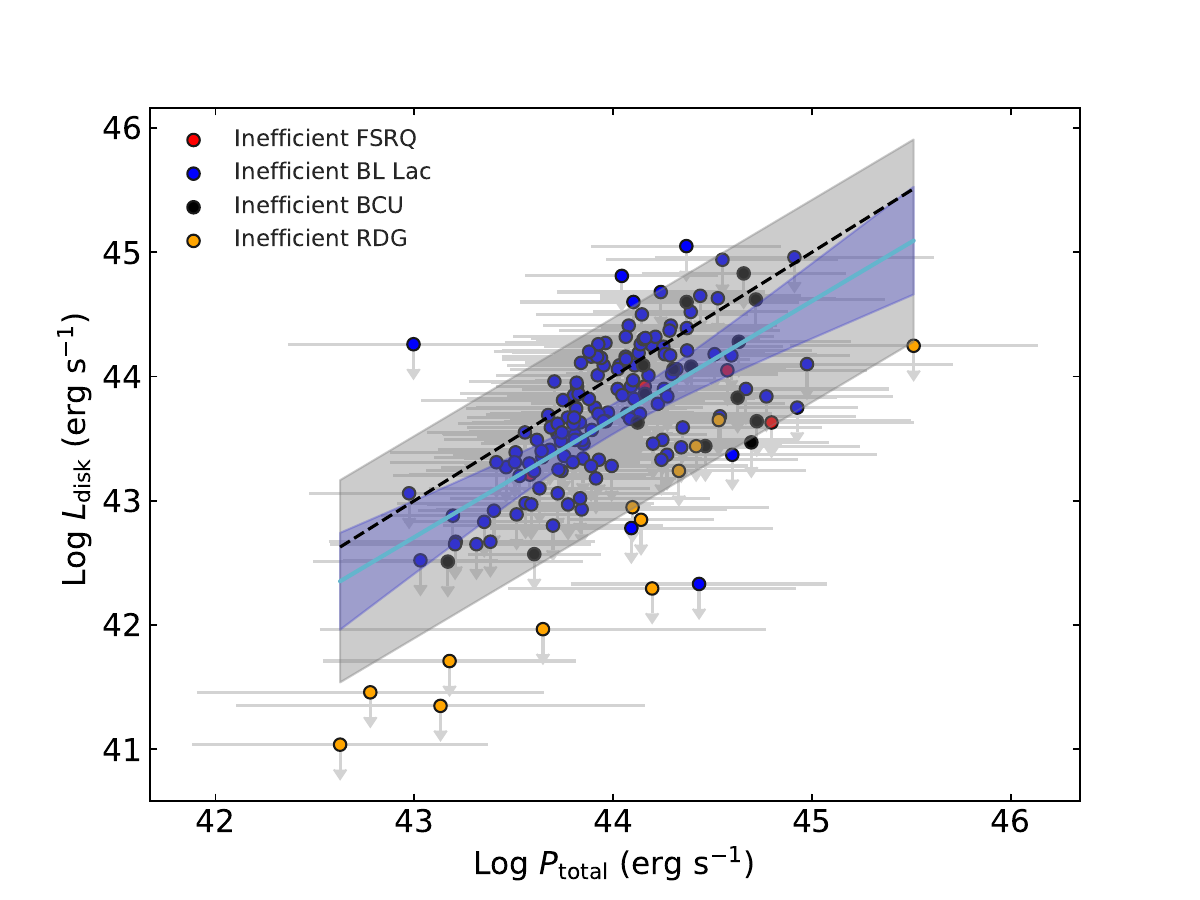}{0.52\textwidth}{(b) Total jet power versus accretion disk luminosity in radiatively inefficient AGNs.}}
\gridline{\fig{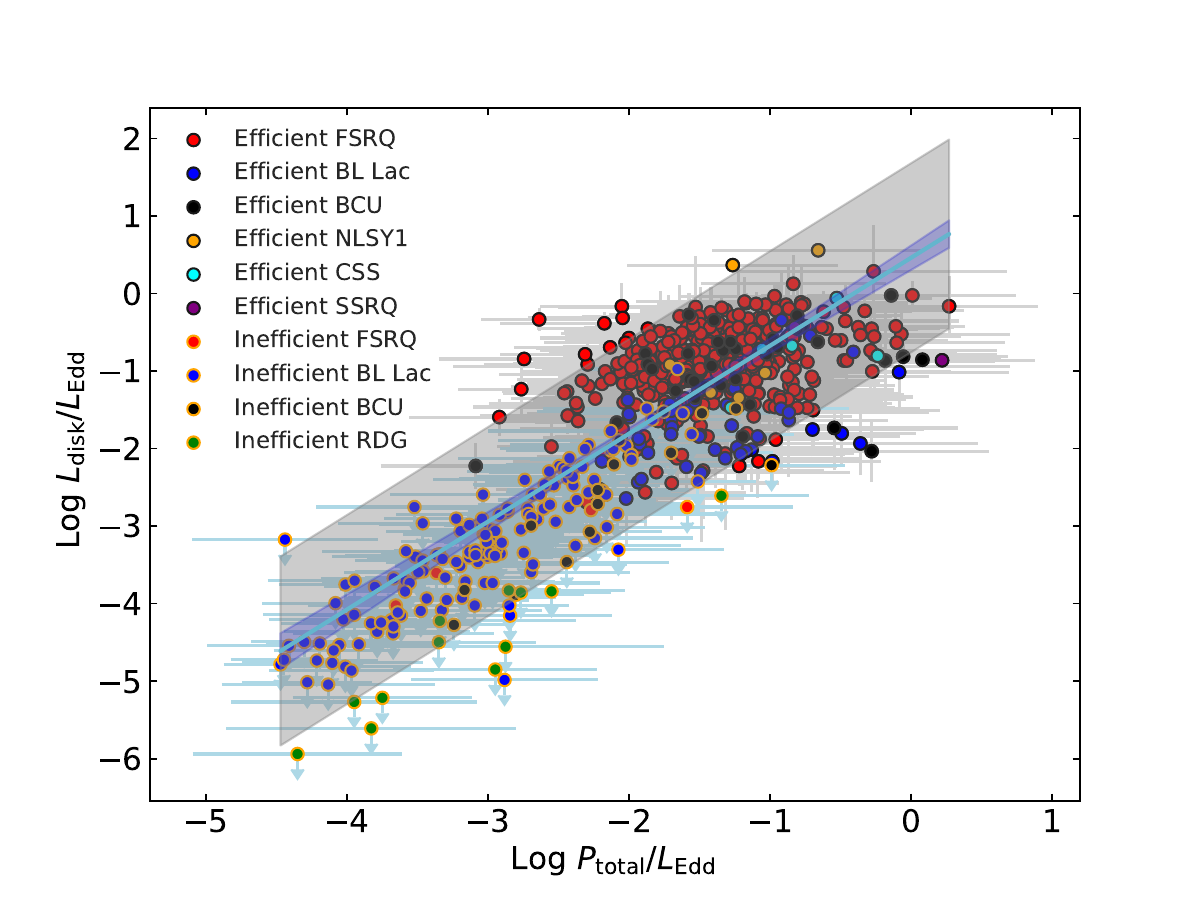}{0.52\textwidth}{(c) jet power (normalized by Eddington luminosity) versus Eddington ratio}
          \fig{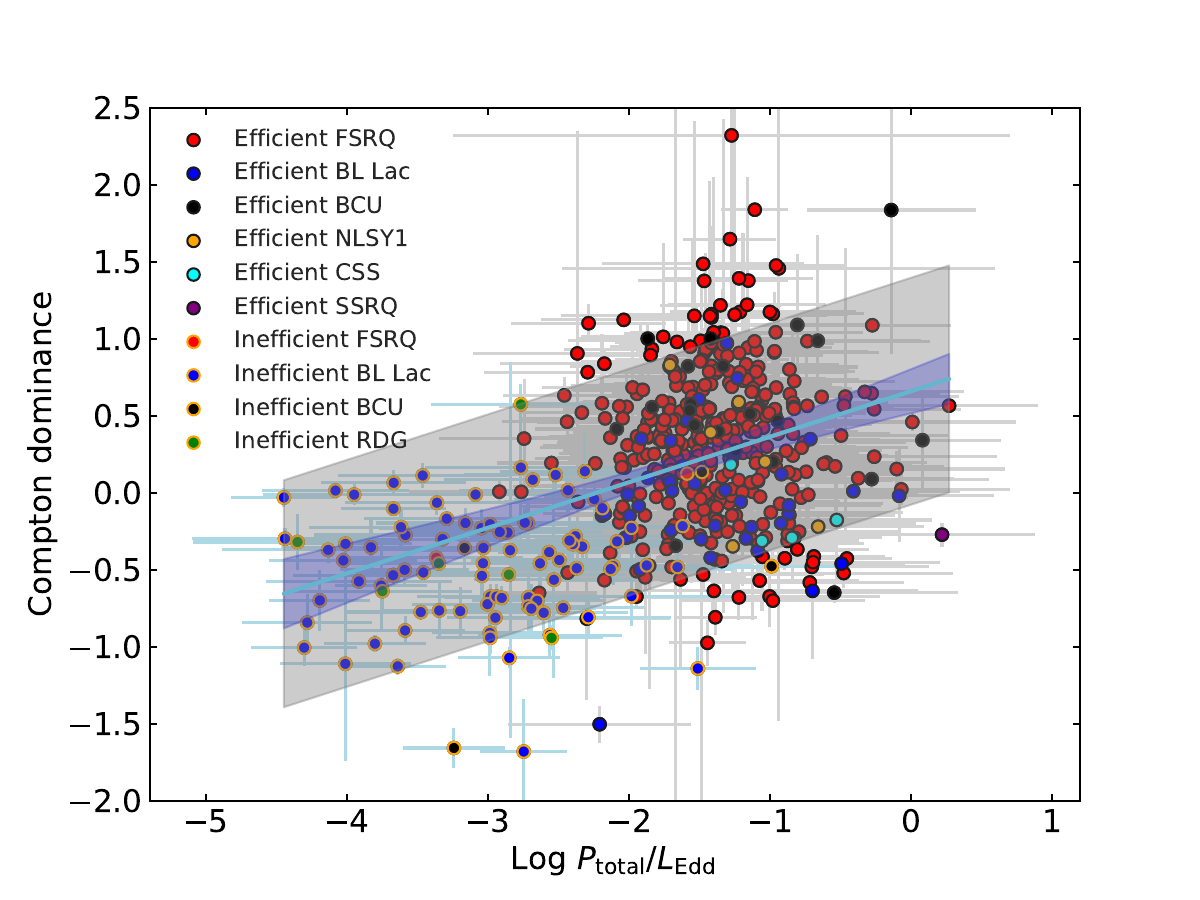}{0.52\textwidth}{(d) jet power (normalized by Eddington luminosity) versus Compton dominance}}
    \caption{Total jet power as a function of accretion disk luminosity, and jet power normalized by Eddington luminosity as a function of Eddington ratio and Compton dominance. The blue lines and shaded areas have the same meaning as in Fig. 8, except that the black dashed line indicates the x = y locus.}
\end{figure*} 

\subsection{The correlations} 

Black hole mass versus total jet power and versus bulk Lorentz factor are presented in Fig. 8. For radiatively efficient AGNs, black hole mass correlates significantly with total jet power (correlation coefficient $r = 0.61$, chance probability $p < 10^{-10}$) and with bulk Lorentz factor ($r = 0.46$, $p < 10^{-10}$) using Ordinary Least Squares (OLS)\footnote{statsmodels.regression.linear\_model.OLS}; $p<0.05$ denotes significance at the 95\% confidence level. The best‑fit linear relation for radiatively efficient AGNs is
\begin{equation}
    \log M_{\rm BH} = 0.68(\pm 0.04) \log P_{\rm total} - 22.27(\pm 1.7).
\end{equation}
For radiatively inefficient AGNs, black hole mass shows almost no correlation with total jet power or bulk Lorentz factor (Fig. 8b and Fig. 8d). Partial‑correlation analyses\footnote{pingouin.partial\_corr} removing common redshift dependence confirm a significant black hole mass–total jet power correlation for radiatively efficient AGNs ($r = 0.42$, $p < 10^{-10}$) and no correlation for radiatively inefficient AGNs ($r = 0.1$, $p = 0.16$).

To test whether these non‑correlations stem from a broad accretion‑rate distribution, we analyze the jet‑power–black‑hole‑mass correlation within accretion‑rate bins (Fig. 9). Four bins are chosen because fewer bins leave each accretion‑rate bin too broad, while more bins result in too few objects per bin. The $r$ values range from 0.2 to 0.51 (weak–to–moderate correlation), with corresponding $p$ values near or below 0.05. Therefore, for radiatively inefficient AGNs, there is a weak-to-moderate correlation between jet power and black hole mass when examined within narrow accretion-rate bins, and a similar relationship exists between bulk Lorentz factor and black hole mass.

There are significant correlations between total jet power and accretion disk luminosity for both radiatively efficient ($r = 0.68$, $p < 10^{-10}$) and radiatively inefficient ($r = 0.63$, $p < 10^{-10}$) AGNs (Fig. 10a-10b). Partial-correlation analyses, removing the common redshift dependence, confirm the significant correlations ($p < 10^{-10}$, $p=3\times10^{-6}$). The best‑fit linear relations are
\begin{equation}
    \log L_{\rm disk} = 1.05(\pm 0.05) \log P_{\rm total} - 1.69(\pm 2.18)
\end{equation}
for radiatively efficient AGNs, and
\begin{equation}
    \log L_{\rm disk} = 0.95(\pm 0.09) \log P_{\rm total} + 1.83(\pm 3.99)
\end{equation}
for radiatively inefficient AGNs. For most radiatively efficient AGNs, jet power does not exceed the accretion‑disk luminosity, whereas for most radiatively inefficient AGNs, jet power exceeds the disk luminosity.

A significant correlation is found between total jet power (normalized by Eddington luminosity) and Eddington ratio across the jetted AGN sample ($r = 0.84$, $p < 10^{-10}$), as shown in Fig. 10c. Partial-correlation analyses removing the common redshift dependence confirm the significant correlations ($r = 0.79$, $p < 10^{-10}$). A similar correlation holds for bulk Lorentz factor normalized by Eddington luminosity versus Eddington ratio ($r = 0.61$, $p < 10^{-10}$), and for bulk Lorentz factor versus Eddington ratio ($r = 0.7$, $p < 10^{-10}$). Because the Eddington ratios for radiatively inefficient AGNs are upper limits, a broken power‑law fit is possible; in this case the slope of the total jet power versus Eddington‑ratio relation for inefficient AGNs is steeper than that obtained from a single power‑law fit.

\citet{Paliya2021} proposed that Compton dominance can serve as an indicator of accretion‑state activity. Jet power normalized by Eddington luminosity as a function of Compton dominance is depicted in Fig. 10d, with significant correlations ($r = 0.44$, $p < 10^{-10}$) also identified. The vast majority of sources lie within the $3\sigma$ dispersion of the best‑fit line. 

A significant correlation exists between total accretion‑disk magnetic field strength and total jet power ($r = 0.4$, $p < 10^{-10}$), as shown in Fig. 11a. However, several sources lie outside the $3\sigma$ dispersion of the best‑fit line. 
The primary reasons may include parameter uncertainties or the inadequacy of a simple power-law model to fully capture their relationship, which might be governed by a more complex underlying mechanism. 

Black hole spin versus jet radiative power is presented in Fig. 11b. When estimating the black hole spin, jet kinetic power is employed. This spin estimate uses as inputs the kinetic power ($P_{\rm kin} = P_{\rm cav}$), and $P_{\rm total}= P_{\rm kin}+P_{\rm r}$. Analyzing spin versus jet kinetic power or total power would introduce x and y autocorrelation. Thus, jet radiative power (a lower limit of total jet power) is used for the correlation analysis. Statistical ﬂuctuations and measurement uncertainties allow sources to have values of $f(j) / f_{\text{max}} > 1$ \citep{Daly2019}. Any $j$ values for which $f(j) / f_{\text{max}} > 1$ are capped at 1. Consequently, in Figure 11b, it can be observed that many sources are truncated at the position $j_{\rm cav}=1.0$ (32\% of the sample), and the diffuse data distribution renders a simple power‑law fit inadequate. The nonparametric Spearman rank test indicates a significant correlation between black hole spin and jet power ($r=0.3$, $p < 10^{-10}$), and a similar significant correlation between black hole spin and bulk Lorentz factor ($r=0.35$, $p < 10^{-10}$).

We also compute black hole spin using $P_{\rm kin}=P_{\rm W}$ with $f=1$ (the smallest jet kinetic power estimator), for which the majority of data do not cluster near 1 (99.7\% of the sampe). A significant correlation ($r=0.46$, $p < 10^{-10}$) between black hole spin and jet radiative power is also found.

Partial‑correlation analyses removing common redshift dependence confirm significant correlations between jet radiative power and black hole spin ($p=5.3\times10^{-9}$, $p=1.2\times10^{-8}$), and between jet radiative power and accretion‑disk magnetic field ($p < 10^{-10}$).

Except for black hole spin and accretion‑disk magnetic field (which lack uncertainties), all correlations are fitted with error weighting, yielding results consistent with the previous correlation analyses. For error-weighted fitting\footnote{statsmodels.regression.linear\_model.WLS}, the weights are defined as $1/{\rm err}^2$, where ${\rm err} = \sqrt{(a \times X{\rm err})^2 + Y_{\rm err}^2}$, with $X_{\rm err}$ and $Y_{\rm err}$ being the errors on the x-axis and y-axis respectively, and $a$ being the non-error weighted fitting coefficient.

\section{Discussions} \label{discussion}

\subsection{Comparative analysis of jet kinetic powers in blazars}

 For the majority of blazars, the jet power estimated via the SED fitting method exceeds that obtained by other methods by one or more ordes of magnitude, even when accounting for uncertainty. In Paper I, we compared SED fitted jet power with jet cavity power and, based on a small sample, also found that SED derived jet power exceeds cavity power. 
In this work, apart from SED‑based estimates, jet powers from other methods are derived from radio or gamma-ray fluxes/luminosities. Possible causes of the discrepancy include: (i) \citet{Ghisellini2014} assumed one proton per radiating electron when fitting the SED, but if the jet contains some electron–positron pairs, the jet kinetic power inferred from SED modeling would be lower; (ii) SED fitting may require more complex jet structures or models; (iii) SED‑derived jet powers likely reflect high‑flux states, whereas cavity‑power or hotspot methods more closely trace long‑timescale averages; (iv) Shocks are often omitted in X-ray cavity jet-power estimates, and using the buoyancy timescale may overestimate ages and thus systematically underestimate jet power \citep{Godfrey2013}.

\begin{figure*}
\centering
\gridline{\fig{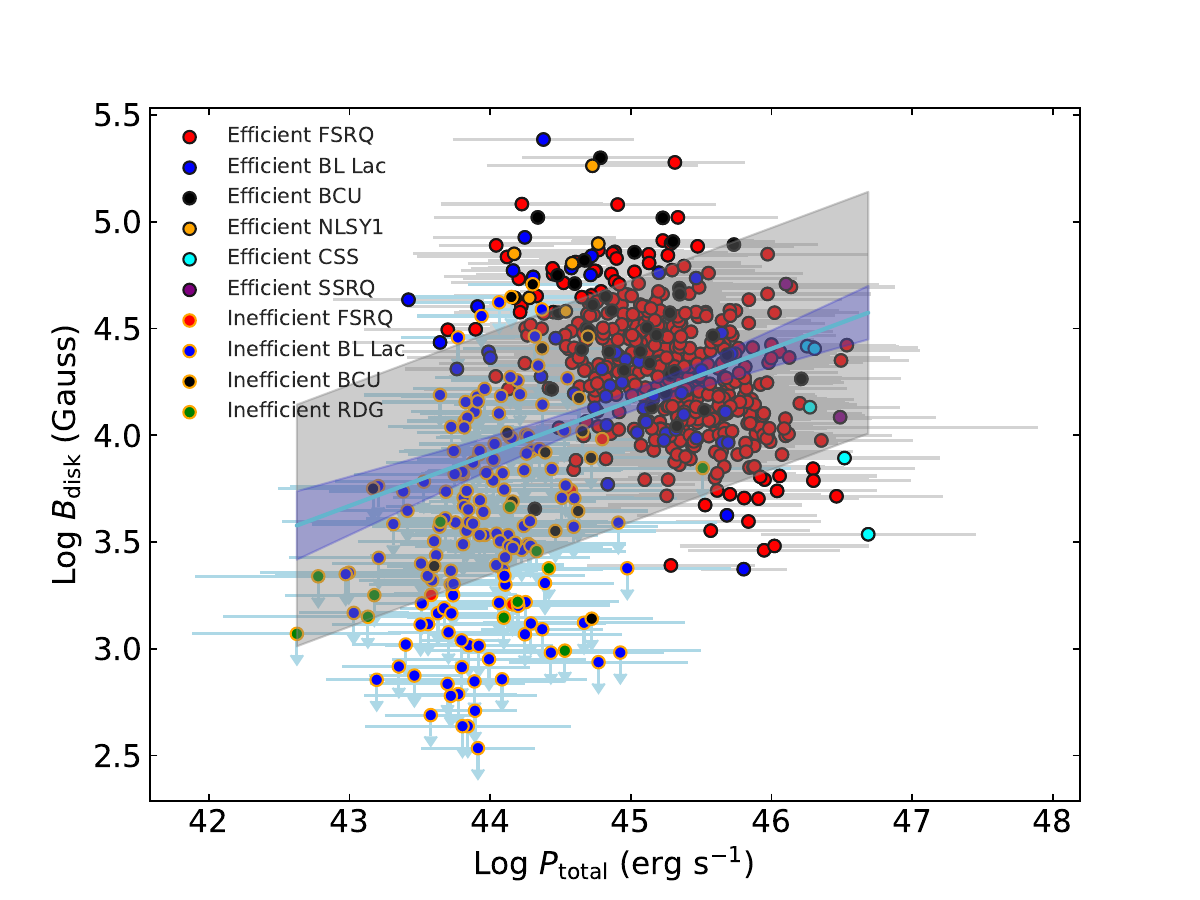}{0.48\textwidth}{(a) Total magnetic field strength of accretion disk versus total jet power} 
          \fig{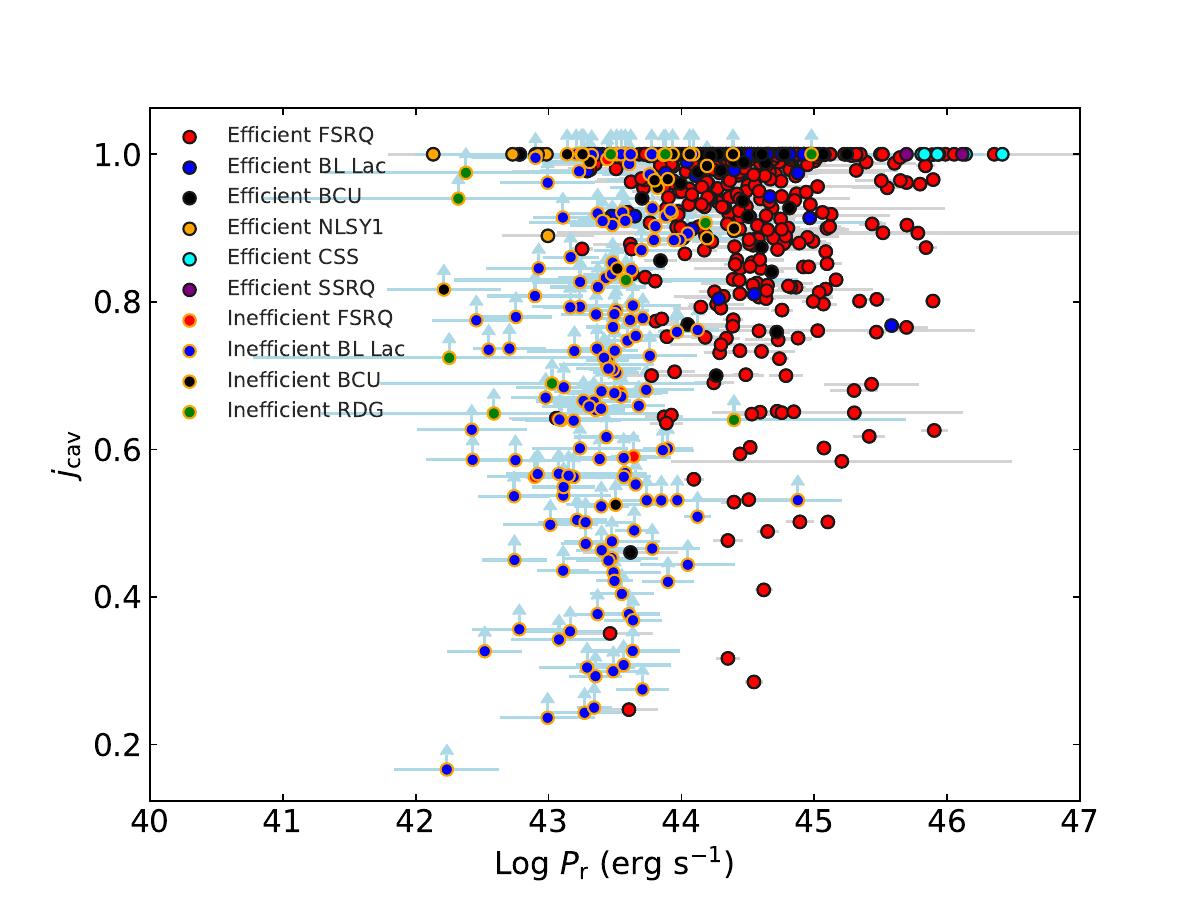}{0.48\textwidth}{(b) Black hole spin versus jet radiative power assuming that $P_{\rm kin} = P_{\rm cav}$ }}
    \caption{Total magnetic field strength of accretion disk as a function of total jet power, and black hole spin as a function of jet radiative power. The blue lines and shaded areas have the same meaning as in Fig. 8.}
\end{figure*} 

\citet{Godfrey2013} measured the kinetic power of 3C FR II sources via a hotspot-based method and compared the $P_{\rm jet}-L_{\rm radio}$ relation with that from FR I X-ray cavity data. Unexpectedly, the FR I and FR II $P_{\rm jet}-L_{\rm radio}$ relations roughly agree. For blazars, jet power estimated via the hotspot method is nearly identical to the cavity power. Therefore, for both radio galaxies and blazars, the hotspot and cavity power methods yield consistent jet kinetic powers.

\subsection{The distributions of the parameters}

\citet{Ghisellini2014} accounted for two jets and external Compton (EC)/synchrotron self-Compton (SSC) processes; thus, relative to the \citet{Ghisellini2012} definition (which we adopt), the radiative power should be scaled by 2×4×(4/3) for EC and 2×16×(4/5) for SSC. The mean jet radiative power of blazars in our 4LAC sample is $10^{44.1}~{\rm erg~s^{-1}}$, versus $10^{45.3}~{\rm erg~s^{-1}}$ in \citet{Ghisellini2014} (2LAC sample). Accounting for definitional offsets of 1.03–1.41 dex, our mean shifts to $10^{45.13–45.51}~{\rm erg~s^{-1}}$, encompassing the \citet{Ghisellini2014} average.

In our 4LAC sample, most sources have jet radiative efficiencies $> 0.03$, with a mean of about 0.2. \citet{Nemmen2012} estimated the jet radiative efficiencies for Fermi 2FGL sources. Our results are broadly consistent with theirs.

For radiatively efficient AGNs, most sources exhibit high spins; for radiatively inefficient AGNs, many likely have high spins, though some may differ. This indicates that black hole spin does not effectively distinguish radiatively efficient from inefficient (jetted) AGNs. Our results also show that black hole spin does not distinguish between subclasses of jetted AGNs. \citet{Volonteri2005} investigated the expected spin distribution of massive black holes and its cosmic evolution within hierarchical galaxy formation, finding that most black holes rotate rapidly at all epochs, implying that high spin is neither necessary nor sufficient to produce a radio‑loud quasar. However, we note that the method to measure black hole spins has some caveats (see Section 4.3).

Spin serves as a fossil record of black hole formation \citep{Reynolds2019}. A population dominated by rapidly spinning black holes points to black hole growth via coherent disk accretion, whereas lower spins favor scenarios involving chaotic, isotropic accretion or mergers of smaller black holes \citep{Reynolds2019,Volonteri2005}. From our results, most radiatively efficient AGNs have high spins, and many radiatively inefficient AGNs likely do as well, though some may differ. Thus, black hole for most radiatively efficient—and many radiatively inefficient—jetted AGNs likely grew via coherent disk accretion, whereas lower‑spin systems likely underwent chaotic, isotropic accretion or mergers of smaller black holes.

Jetted AGNs can be classified by their accretion rates \citep[e.g.,][]{Ghisellini2001,Ghisellini2011,Sbarrato2012,Xiong2014,Padovani2017}. In radiatively efficient AGNs, the total magnetic field strength of the accretion disk spans $10^{3.3}-10^{5.5}$ Gauss, with a mean of $10^{4.32\pm0.01}$ Gauss, indicating a strong magnetic field strength. The magnetic field strength of the accretion disk in most radiatively efficient AGNs exceeds that in radiatively inefficient AGNs. 
We propose distinguishing radiatively efficient and inefficient AGNs by the differing magnetic field strengths of their accretion disks. Accounting for measurement uncertainties and upper limits from inefficient AGNs, the dividing magnetic field is $\approx10^{3.9}$ Gauss. Possible caveats: the dividing line is a smooth rather than sharp transition, and magnetic field strengths for inefficient AGNs are upper limits.

\subsection{Outflow-based methods for estimating black hole spin}

The key assumption of the outflow method is that the jet or outflow is at least partially powered by black hole spin. For a Fermi blazar sample, \citet{Chen2021} reported that the powers of relativistic jets depend on the spins of accreting SMBHs. Relativistic jet power was found to exceed the accretion‑disk luminosity in blazars \citep{Ghisellini2014,Chen2018}, indicating that jet formation is likely dominated by the Blandford–Znajek mechanism. Other studies likewise suggest that the jet or outflow may be at least partially powered by black hole spin \citep[e.g.,][]{Xiong2014,Zhang2022,Zhang2014,Xiao2022,Chen2023,Wu2011,Chai2012}. Furthermore, our results also support that jets are at least partially powered by black hole spin (see section 4.4). Therefore, in this study, estimating black hole spin via the outflow method is feasible. Daly et al. emphasize that the method of outflow does not depend upon a specific jet production model or a specific accretion disk model \citep{Daly2024,Daly2019}.

Due to statistical fluctuations and measurement uncertainties, some sources may exhibit values of $f(j)/f_{\text{max}}$ greater than 1 \citep{Daly2019}. Numerical simulations suggest that the conversion from the spin function to spin can change depending on the details of the model for the black hole system \citep[e.g.,][]{Tchekhovskoy2010,Yuan2014}, and thus this may be a source of uncertainty \citep{Daly2019}. 

\citet{Daly2019} applied outflow methods to estimate black hole spin and accretion‑disk magnetic fields in LINERs, FR II, and compact radio sources, finding no link between source type and spin among systems with collimated outflows. In our sample, black hole spin does not distinguish between jetted AGN subclasses. Thus, for both blazar and non‑blazar jetted AGNs, black hole spin does not determine AGN subtype. The spins in the \citet{Daly2019} sample are relatively high, typically ranging from about 0.6 to 1. 
Our sample shows a consistent distribution, though spins below 0.6 are possible for radiatively inefficient sources.

Caveats when using this scaling to estimate spin: (1) jet power is difficult to measure accurately; (2) extended 300 MHz–based estimates have large uncertainties; (3) results rely on an inner accretion‑flow model; (4) independent confirmation with other spin methods is needed.

\subsection{The correlations}

We conduct OLS fits, redshift‑controlled partial correlations and error‑weighted analyses, obtaining consistent correlation results.
\subsubsection{Black hole mass versus total jet power and versus bulk
Lorentz factor}

The correlation between black hole mass and jet power has been widely investigated. In theory, \citet{Heinz2003} derived how jet flux scales with black hole mass in core‑dominated systems: $F_\nu \propto M^{17/12}$ for both standard and radiatively inefficient accretion. Within the BZ framework, the jet power is predicted to depend on the black hole mass, its spin and the magnetic field at the event horizon \citep{Blandford1977,Ghisellini2014}. From our results, jet power correlates significantly with black hole mass in radiatively efficient AGNs, while in radiatively inefficient AGNs a weak‑to‑moderate correlation emerges within narrow accretion‑rate bins.

Observationally, a significant correlation between jet power and black hole mass has been established \citep[e.g.,][]{Merloni2003,Liu2006,Xiong2014,Wang2004}. 
In BL Lac objects, studies have reported only a weak correlation between black hole mass and jet power \citep[e.g.,][]{Chen2023b,Zhang2012}. Results from \citet{Fan2019} indicate that jet power depends not only on black hole mass but also on other physical properties. Compared with prior studies, we use one of the largest jetted‑AGN samples spanning all subclasses, classify them into radiatively efficient and inefficient categories, and consider total jet power. Jet power correlates significantly with black hole mass in radiatively efficient AGNs, whereas radiatively inefficient AGNs show little to no correlation when not restricted to narrow accretion‑rate bins.
The upper limits of the Eddington ratio for radiatively inefficient AGNs span five orders of magnitude (Fig. 10c), and weak-to-moderate correlations emerge when examined within narrow accretion-rate bins. Therefore, the lack of correlation between black hole mass and jet power may result from the wide spread in Eddington ratios.

Compared with black hole mass versus jet power, the relation between black hole mass and bulk Lorentz factor is less explored. \citet{Chai2012} found a significant correlation between jet power and black hole mass, but no correlation with the Eddington ratio. If more massive black holes tend to have higher spins, this implies that jet power is likely governed by black hole spin. \citet{Xiong2014b} reported similar results for FSRQ samples. From our sample, the bulk Lorentz factor correlates significantly with black hole mass in radiatively efficient AGNs. For radiatively inefficient AGNs, a weak‑to‑moderate correlation emerges within narrow accretion‑rate bins.

\subsubsection{Dependence of total jet power on accretion‑disk luminosity and of Eddington-luminosity‑normalized jet power on Eddington ratio}

If relativistic jets are powered, at least initially, by Poynting flux, the Blandford–Znajek power can be written as \citep{Ghisellini2006}:
\begin{equation}
    L_{\mathrm{BZ}} \sim \left(\frac{a}{m}\right)^{2} \frac{R_{s}^{3}}{R^{2} H} \frac{\varepsilon_{B} \, L_{\mathrm{disk}}}{\eta \, \beta_{R}}
\end{equation}
where $a/m$ is the specific black hole angular momentum ($\sim 1$ for maximally rotating black holes); $R_{s}$ the Schwarzschild radius; $H$ the disk thickness; $R$ the radius; $\varepsilon_{B}$ the fraction of the available gravitational energy; $\eta$ the accretion efficiency; $L_{\mathrm{disk}}$ the observed disk luminosity; $\beta_{R}c$ the radial infalling velocity. 
The maximum jet power is obtained by setting $R \approx H \approx R_{s}$ and taking $a/m$, $\varepsilon_{B}$, and $\beta_{R}$ to unity; in this case, the expression simplifies to its maximal form:
\begin{equation}
    L_{\mathrm{BZ,max}} \sim \frac{L_{\mathrm{disk}}}{\eta}.
\end{equation}

From our results, There are significant correlations between total jet power and accretion disk luminosity for both radiatively efficient and radiatively inefficient AGNs. The near‑unity slope between jet power and disk luminosity supports jets powered by energy extraction from rapidly spinning black holes through accretion‑disk magnetic fields. A near‑unity slope was also found for a Fermi blazar sample by \citet{Xiong2014}.

Beyond the jet–black hole mass relation, the jet–accretion connection has also been extensively studied \citep[e.g.,][]{Rawlings1991,Cao1999,Wang2004,Maraschi2003,Liu2006,Ghisellini2010,Gu2009,Paliya2017,Chen2023b}. Here, all jetted AGN subclasses are included, especially young CSS sources and low‑accretion‑rate radio galaxies, and blazar beaming effects are excluded. Our results reveal significant correlations between the Eddington ratio and both the jet power and the bulk Lorentz factor (normalized by Eddington luminosity), supporting the jet–accretion connection in jetted AGNs. In the BZ framework, jets require accretion to provide the magnetized plasma and ﬁeld to thread the ergosphere of the black hole and a spinning black hole to provide an energy source to power and accelerate the jets \citep{Blandford1977,Daly2019}.

Note that Eddington ratios for radiatively inefficient AGNs are upper limits, and a steeper slope is therefore possible for this population. \citet{Sbarrato2014} used a broken power law to fit the luminosity of the BLR (in Eddington units) versus the radio luminosity (in Eddington units). Below the break value, jetted AGNs are radiatively inefficient and the slope steepens.

A significant correlation is found between Compton dominance (an accretion‑state indicator) and jet power normalized by Eddington luminosity. This helps mitigate apparent correlation between jet power and Eddington ratio driven by upper limits in radiatively inefficient AGNs.

\subsubsection{Jet power versus total accretion-disk magnetic field strength and black hole spin}

The formation mechanism of highly relativistic jets in AGNs is a major open question in astrophysics. Jets are thought to form near the central black hole, powered by the extraction of energy from the black hole's spin (BZ) or its accretion disk (BP) via magnetic fields. Because the accretion of matter sustains these magnetic fields, a direct relationship between accretion power and jet power is theoretically predicted \citep{Maraschi2003}. If the total accretion‑disk magnetic field is linked to these magnetic fields, a correlation between jet power and total accretion‑disk magnetic field should be expected. In our sample, we find a significant correlation between total accretion‑disk magnetic field strength and total jet power, consistent with this theoretical prediction.

The nonparametric Spearman test shows significant correlations between black hole spin and jet radiative power, and between spin and bulk Lorentz factor, supporting that black hole spin energy powers and accelerates jets. These significant correlations also support the key hypothesis of the outflow method.

\section{Conclusions} \label{sec:Conclusion}

A catalog of key parameters for powerful jet–accretion disk–black hole systems is presented (Tables 1–3). The catalog includes kinetic and radiative jet powers, jet radiative efficiency, bulk Lorentz factor, black hole spin, total accretion disk magnetic field, and Compton dominance. This constitutes one of the largest jetted AGN samples with such measurements and includes all jetted AGN subclasses. The main conclusions are as follows.

(i) Utilizing low-frequency extended radio luminosity to estimate jet kinetic power via the cavity power relation, we compare these results with estimates from SED fitting, hotspot measurements, and other scaling relations derived from radio or gamma‑ray fluxes/luminosities. For most of blazars, SED‑derived jet powers exceed those from other methods by one or more orders of magnitude, even after accounting for uncertainties. Possible causes of the discrepancy are discussed.

(ii) The mean jet kinetic powers for gamma‑ray–emitting jetted AGNs rank as: ${\rm CSS>SSRQ>FSRQ>BCU>NLSY1>Other~AGN}$ ${\rm >BL~Lac>RDG}$. However, the numbers of RDGs, CSSs, other AGNs, NLSY1s, and SSRQs in the present sample
  are very small and therefore for those sources this result should be treated with caution.

Most sources have jet radiative efficiencies $> 0.03$, with a mean of about 0.2, indicating that jets in gamma‑ray–emitting AGNs are dominated by kinetic rather than radiative processes. 

(iii) For radiatively efficient AGNs, the majority of sources exhibit black hole spin values exceeding 0.9, with an average spin of $0.929 \pm 0.005$. In contrast, for radiatively inefficient AGNs, the average lower limit of the black hole spin is 0.71; it is highly probable that many of these sources have high spins ($>0.9$), though some may exhibit different spin states. Our results show that black hole spin does not effectively distinguish between subclasses of jetted AGNs. Because the black hole spins used in this study were inferred from scaling relations, confirmation using other spin estimation methods is important.

Black hole for most radiatively efficient jetted AGNs and many radiatively inefficient ones likely grew their masses via coherent disk accretion, while lower‑spin jetted AGNs likely underwent chaotic, isotropic accretion or mergers of smaller holes.

(iv) In radiatively efficient AGNs, the total magnetic field strength of the accretion disk spans $10^{3.3}-10^{5.5}$ Gauss, with a mean of $10^{4.32\pm0.01}$ Gauss, indicating a strong magnetic field strength. We propose distinguishing radiatively efficient and inefficient AGNs by the differing magnetic field strengths of their accretion disks. Accounting for measurement uncertainties and upper limits from inefficient AGNs, the dividing magnetic field is $\approx10^{3.9}$ Gauss. 

(v) Jet power correlates significantly with black hole mass in radiatively efficient AGNs, while in radiatively inefficient AGNs a weak‑to‑moderate correlation emerges within narrow accretion‑rate bins. A similar pattern holds for bulk Lorentz factor versus black hole mass. Note possible biases in the derived physical quantities due to different methods that have been used to derive them.

(vi) There are significant correlations between total jet power and accretion disk luminosity for both radiatively efficient and radiatively inefficient AGNs. The near‑unity slope between jet power and disk luminosity supports jets powered by energy extraction from rapidly spinning black holes through accretion‑disk magnetic fields.

(vii) Significant correlations are found between Eddington ratio and both jet power and bulk Lorentz factor (normalized by Eddington luminosity), and between Compton
dominance (an accretion-state indicator) and jet power normalized by Eddington luminosity, supporting the jet–accretion connection in jetted AGNs.

(viii) We report a significant correlation between total accretion‑disk magnetic field strength and total jet power, consistent with theoretical prediction. The nonparametric Spearman test shows significant correlations between black hole spin and jet radiative power, and between spin and bulk Lorentz factor, supporting that black hole spin energy powers and accelerates jets.

The correlations discussed in (vi)--(viii) involve quantities that share a common dependence on redshift. Nevertheless, these correlations persist after controlling for redshift in partial-correlation analyses.

\begin{acknowledgments}
This work is supported by the National Key R\&D Program (2023YFE0101200), the National Natural Science Foundation of China (Grant Nos. 12473020, 12025303 and 12393813), the Yunnan Province Youth Top Talent Project (Grant No. YNWR-QNBJ-2020-116), the Yunnan Revitalization Talent Support Program (YunLing Scholar Project), the Yunnan Province Foundation (Grant No. 202301AU070160), the CAS ``Light of West China" Program. Y.C. is supported by the Training Program for Talents in Xingdian, Yunnan Province (Grant No. 2081450001) and the National Natural Science Foundation of China (Grant No. 12203028). M.G. is supported by the National Science Foundation of China (grant No. 12473019), the National SKA Program of China (grant No. 2022SKA0120102), the Shanghai Pilot Program for Basic Research-Chinese Academy of Science, Shanghai Branch (grant No. JCYJ-SHFY-2021-013), and the China Manned Space Project with No. CMS-CSST-2025-A07.

This research has made use of the NASA/IPAC Extragalactic Database, which is funded by the National Aeronautics and Space Administration and operated by the California Institute of Technology.
\end{acknowledgments}

\software{NumPy \citep{Harris2020}, Matplotlib \citep{Hunter2007}, Astropy \citep{Astropy2022,Astropy2013,Astropy2018}, SciPy \citep{Virtanen2020}, Pandas \citep{mckinney2010,reback2020}, Statsmodels \citep{seabold2010}, Seaborn \citep{Waskom2021}, TOPCAT \citep{Taylor2011}, Pingouin \citep{Vallat2018}
          }



\bibliography{sample7}{}
\bibliographystyle{aasjournal}



\end{document}